\documentclass[]{raa_rb}
\usepackage{graphicx,times}
\usepackage{natbib}
\usepackage{psfrag}
\usepackage{amsmath}
\usepackage{booktabs,longtable}
\usepackage{pdflscape}

\begin{document}

\title{New Open Cluster Candidates Discovered in the XSTPS-GAC Survey}
   \volnopage{Vol.0 (200x) No.0, 000--000}      %%preserved for Editor. DOn't remove!
   \setcounter{page}{1}          %%starting page, preserved for Editor. DOn't remove!

   \author{Jincheng Guo
      \inst{1,2,3,8}
   \and Huawei Zhang
      \inst{1,2}
   \and Huihua Zhang
      \inst{1,2}
    \and Xiaowei Liu
      \inst{4,1,2}
     \and Haibo Yuan 
      \inst{5}
     \and Yang Huang
       \inst{4,1,2,8}
      \and Song Wang
       \inst{3}
      \and Li Chen
       \inst{6}
      \and Haibin Zhao
       \inst{7}
      \and Bingqiu Chen
       \inst{4}
      \and Maosheng Xiang
       \inst{3,8}
      \and Zhijia Tian 
        \inst{1,2,8}
      \and Zhiying Huo
        \inst{3}  
       \and Chun Wang
        \inst{1,2}
   }
   \institute{$^{1}$Department of Astronomy, Peking University, Beijing 100871, P. R. China;{\it jincheng.guo@pku.edu.cn;zhanghw@pku.edu.cn}\newline
   $^{2}$Kavli Institute for Astronomy and Astrophysics, Peking University, Beijing 100871, P. R. China \\
   $^{3}$Key Laboratory of Optical Astronomy, National Astronomical Observatories, Chinese Academy of Sciences, Beijing 100012, China\\
   $^{4}$South-Western Institute for Astronomy Research, Yunnan University, Kunming 650500, China\\
   $^{5}$Department of Astronomy, Beijing Normal University, Beijing 100875, China\\
   $^{6}$Shanghai Astronomical Observatory, Shanghai 200030, P. R. China\\
   $^{7}$Purple Mountain Observatory, Chinese Academy of Sciences, Nanjing 210008, P. R. China\\
   $^{8}$LAMOST Fellow
   }

\abstract
{The Xuyi Schmidt Telescope Photometric Survey of the Galactic Anti-center (XSTPS-GAC) is a photometric sky survey that covers nearly 6\,000 deg$^{\rm 2}$ towards Galactic anti-center in $g, r, i$ bands. Half of its survey field locates on the Galactic Anti-center disk, which makes XSTPS-GAC highly suitable for searching new open clusters in the GAC region. In this paper, we report new open cluster candidates discovered in this survey, as well as properties of these open cluster candidates, such as age, distance and reddening, derived by isochrone fitting in the color-magnitude diagram (CMD).  These open cluster candidates are stellar density peaks detected in the star density maps by applying the method from \cite{Koposov2008}.  Each candidate is inspected on its true color image composed from XSTPS-GAC three band images. Then its CMD is checked, in order to identify whether the central region stars have a clear isochrone-like trend differing from the background stars. The parameters derived from isochrone fitting for these candidates are mainly based on three band photometry of XSTPS-GAC. Meanwhile, when these new candidates are able to be seen clearly on 2MASS, their parameters are also derived based on 2MASS ($J-H,J$) CMD. Finally, there are 320 known open clusters rediscovered and 24 new open cluster candidates discovered in this work. Further more, the parameters of these new candidates, as well as another 11 known recovered open clusters, are properly determined for the first time. 
}

   \authorrunning{Guo et al.}       
   \titlerunning{New Open Cluster Candidates of XSTPS-GAC } 

   \maketitle

\section{INTRODUCTION}

It is commonly accepted that star clusters are the building blocks of galaxy.  They are the major star birth environment \citep{Lada2003,deWit2005}. Thus, star clusters play an important role in the investigations of star formation, stellar evolution, and can be used as tracers of structure and evolution of their host galaxy.  Since star clusters usually contain tens to thousands of stars with single-age and single-metallicity, their determined fundamental parameters, such as age, distance and reddening, can be more accurate and reliable.  Meanwhile, star cluster is a natural laboratory to test the validation of spectroscopic methods determined stellar parameters like $T_{\rm eff}$, $log~g$ and $metallicity$, as in \cite{xiang2015}.

There are 2\,167 open clusters in the most recent version(Jan 28, 2016) of DAML02 catalog \citep{Dias2002}. These clusters are collected from various literature, therefore, they are based on different observations and their cluster parameters are derived by different methods.  \cite{Kharchenko2012, Kharchenko2013} collects 3\,784 clusters from literature in their MWSC catalog, including open clusters, globular clusters, embedded clusters and star associations. Homogeneous cluster parameters have been determined for 3\,006 clusters, using 2MASS \citep{Skrutskie2006} photometry combined with position and proper motions from PPMXL \citep{Roser2010}.  However, the number of known star clusters in the Milky Way is still much less than the number of theories predicted, which is more than 20,000 \citep{Zwart2010, Bonatto2006}. 

Many large area sky surveys have been used to find clusters and derive cluster parameters. Among those surveys, 2MASS, SDSS, DSS are the most useful ones, adopted by the work of  \cite{Bonatto2006}, \cite{Bica2006}, \cite{An2009} and  \cite{Kronberger2006}. Recently, \cite{Borissova2011} used data from VISTA Variables in the Via Lactea survey \citep[VVV,][]{Minniti2010} to find new open clusters in the Galactic center region, which suffers serious dust extinction. VVV was also used to improve the accuracy of cluster parameters \citep{Majaess2012}. On the other hand, large astrometric catalogs, such as UCAC4 \citep{Zacharias2013}, GSC2.3 \citep{Lasker2008}, PPMXL\citep{Roser2010}, compiled from various sky surveys could provide position and proper motion of high precision, while proper motion is the most important information to identify cluster member stars \citep{Kharchenko2005}.

In the early days, bright open clusters were discovered by eye. And for some large and loose clusters, they were found by proper motion analysis. Nowadays, new effective methods are developed to find more clusters. For example, more and more clusters are discovered by automatic methods.  \cite{FSR2007} found 1\,021 clusters based on stellar density maps from 2MASS point source catalog, which greatly increased the sample of known clusters.  \cite{Koposov2008} constructed a 2-D filter based on the difference of two Gaussian profiles. By convolving the stellar density map with this filter, cluster candidates were found more effectively.

The Xuyi Schmidt Telescope Photometric Survey of the Galactic Anti-center (XSTPS-GAC) is a large sky survey in SDSS $g,r,i$ bands.  It is the photometric part of the Digital Sky Survey of the Galactic Anti-center (DSS-GAC, Liu et al. 2015), and DSS-GAC is a photometric and spectroscopic sky survey \citep{xiaowei2013}. A significant part of LAMOST survey \citep{gang2012} is the spectroscopic part of DSS-GAC. XSTPS-GAC covers a continuous sky area of over 6\,000 $\rm deg^2$ towards Galactic Anti-center (GAC). Nearly half of XSTPS-GAC's survey field locates on the GAC disk, which makes XSTPS-GAC highly suitable for searching new open clusters in the GAC region. Comparing to 2MASS, which currently is the main catalog used to find open clusters, XSTPS-GAC reaches a deeper limiting magnitude of $r \approx$ 19.0 mag. This fact will lead to more observations of distant faint stars, which would also suffer heavier dust extinction.  Therefore, we except to discovery new open clusters from XSTPS-GAC photometric data.

The paper is organized as follows.  In Section 2, we briefly present the
observations and data reduction of XSTPS-GAC.  The method to find cluster
candidates are described in Section 3.  We present the result and discussion on the
interesting clusters in Section 4.  Finally, we give the conclusions in Section
5.

\section{Observation and Data Reduction}
\label{sec2}
The XSTPS-GAC started collecting data in October 2009 and completed in March 2011, using the Xuyi Schmidt telescope that operated by Purple Mountain Observatory, Chinese Academy of Sciences. The telescope has a 120\,cm primary mirror and a 104\,cm correcting plate.  A 4096$\times$4096 CCD camera is equipped in the telescope, with a 1.94$^{\rm o}$$\times$1.94$^{\rm o}$ effective field-of-view and a pixel scale of 1.705 arcsec. Seeing at the site can reach up to 1$^{''}$ under the best air condition.  During XSTPS-GAC operation period, the mean atmospheric extinction coefficients are 0.69, 0.55 and 0.38 mag/airmass for the $g,r,i$ bands, respectively. The typical night sky brightness could reach to 21.7, 20.8 and 20.0 mag $\cdot$ arcsec$^{\rm -2}$ for these three bands \citep{zhang2013}.

XSTPS-GAC collected more than 20\,000 CCD images in $g$, $r$, $i$ bands,
covering a continuous GAC area from 3\,$\rm h$ to 9\,$\rm h$ in right ascension,
from -10$^{\rm o}$ to +60$^{\rm o}$ in declination, plus an extended area of
about 900 deg$^2$ towards M31/M33.  We designed 39 observation stripes to cover
the whole GAC region.  Each stripe is consisted of a group of continuous fields under a
fixed declination, and each field has a 0.95 deg area, about half of a field, overlapping with its
adjacent fields in the same stripe. As a result, a single source in XSTPS-GAC planed region
has been exposed at least two times for each band. The observations were carried out by adopting 90\,s as exposure time, using only one filter for each night and observing two stripes together.
This observing strategy not only enabled the observation to match the sky moving speed well, but also ensured the whole night observation's consistency in the sense of airmass.

The data reduction was carefully carried out, including overscan trimming, bias
correction, flat-fielding, source detection and subtraction in typical
$DAOPHOT$\citep{Stetson1987b} steps for CCD images.  Flat-fielding is
accomplished by applying super-sky-flat (SSF) method, which assumed the sky
background is homogeneous. The SSF field of each night was subtracted from
hundreds of raw observed fields in the same filter. Comparing with dome flat
or twilight flat, which maybe suffer from streaks, non-linearity and
large-scale inhomogeneities on the field, our SSF field correction accuracy
could reach 1\%.  We also divided one CCD image into 5$\times$5 sub-fields, in order to
construct star point-spread-function (PSF) more precisely, which ensures robust
PSF flux measurement for all detected stars. Based on overlapping large area
between adjacent fields, the photometry calibration was done in ubercal calibration
manner, same way as SDSS \citep{Padmanabhan:2008}. The typical photometric accuracy could
reach up to 0.03 mag (Yuan et al., 2017, in preparation). By adopting PPMXL as the
astrometric reference, the astrometric accuracy could reach 70 mas for
stars brighter than 17.0 mag, and  better than 200 mas for stars at the faint end ($\sim 19.0$ mag) of
the survey. \citep{Zhang2014}.

\section{Cluster search and analysis}
\subsection{Candidates Search}
We applied method from \cite{Koposov2008} to detect high stellar density regions in XSTPS-GAC. 
First, stellar density map was constructed based on each CCD image,  
which could improve the speed of convolution in the following process. Stars are counted in the grid size of 10$\times$10 CCD pixels, namely $\rm 17.5\times17.5\,arcsec^{2}$. The selection of this grid size is a result of considering small scale density variation will be neglected if a larger grid size was chosen and there will be many zero counts if a smaller grid size was chosen. 
In addition, we selected stars that have $r$ and $i$ band measurements at least, since $g$ band suffers from relatively heavier dust extinction, more stars are able to be detected in $r$ and $i$ band. Also we constrained the photometric error to be less than 0.1 mag, ensuring the photometric quality. The filter from \cite{Koposov2008} was adopted in this work, shown in Equation (\ref{eq:filter}). 

\begin{equation}
\label{eq:filter}
\begin{aligned}
& S(ra,dec)  =  \\
& \sqrt{4\pi }{\sigma}_{1}\frac{M(ra,dec)*(G(ra,dec,{\sigma}_{1})-G(ra,dec,{\sigma}_{2}))}{\sqrt{M(ra,dec)*G(ra,dec,{\sigma}_{2})}}\\
\end{aligned} 
\end{equation}

Where $M(ra,dec)$ represents the star number density map, while $G(ra,dec,\sigma)$ is the circular 2D Gaussian with unity integral and width of $\sigma$. 

\begin{figure*}
 \begin{center}
 \includegraphics[width=150mm]{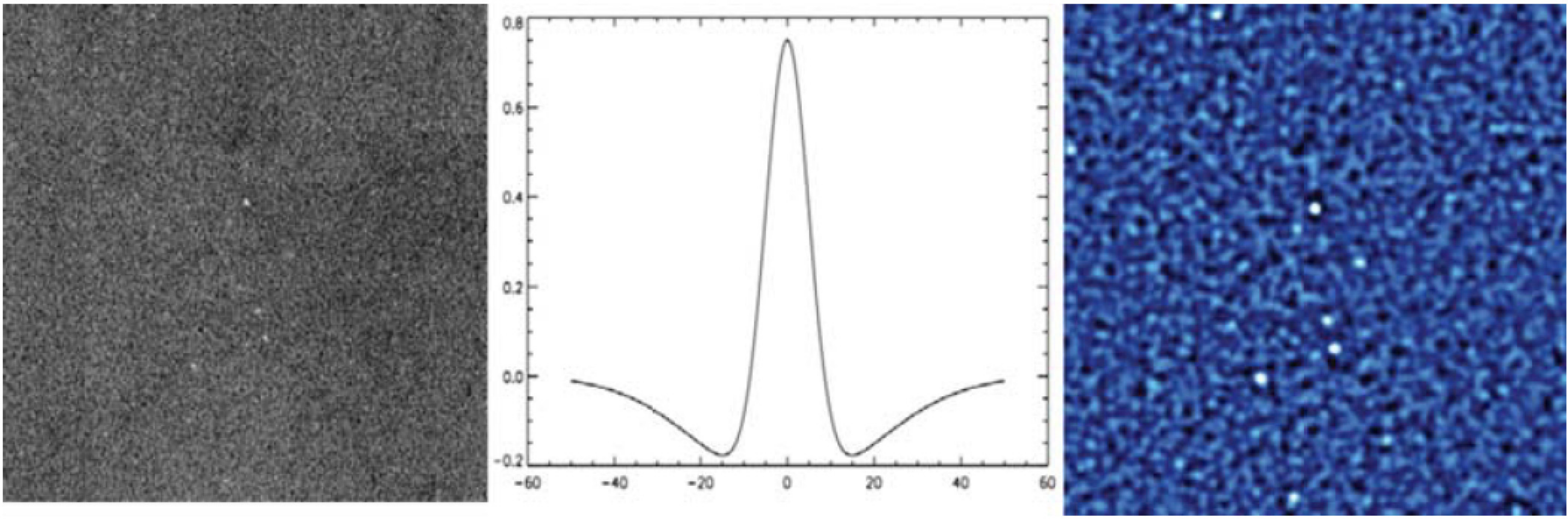}
 \caption{The $Koposov$ method adopted to search for the over-density area. The left panel is the stellar number density map based on data from XSTPS-GAC. Constructed $Koposov$ filter is shown in the middle panel. The right panel shows the stellar number density map after convolution. The over-density area is more evident.} 
 \label{fig0}
 \end{center}
 \end{figure*}

The convolved map $S(ra,dec)$ shows the deviation of $M(ra,dec)*G(ra,dec,{\sigma}_{1})$ above the background estimation, namely $M(ra,dec)*G(ra,dec,{\sigma}_{2})$. The over-density detections on this map can be easily acquired by setting $S(ra,dec) > 5$, which represents over-density statistically significance above 5$\sigma$. The 1D slice of 2D filter with convolution for $\sigma_{1} = 3^{\prime}$ and $\sigma_{2} = 6^{\prime}$ is shown in Fig. \ref{fig0}, similar filter from \cite{Koposov2008}. In this work, $S(ra,dec)>4.5$, $\sigma_{1} = 2^{\prime}$ and $\sigma_{2} = 6^{\prime}$ were used, in order to include more small-sized clusters.

Due to the fact that half of each XSTPS-GAC plate is overlapped with adjacent ones, the true density peaks will be detected at least two times. In this sense, density peaks produced by edge effect can be easily removed. However, after the removal of edge effect, we still found lots of density peaks in relatively high galactic latitude regions. Some of them could be real star clusters, but significant number of these detections should be caused by field star number density fluctuations. According to \cite{Gilmore1983}, star density decreases along galactic latitude in a power-law manner. Therefore, in the low latitude dense region populated by field stars, the density variation feature of true star clusters will be less significant. While in the relatively high latitude sparse region, many over-density detections are in fact due to the magnification of star density fluctuation. In order to find more true cluster over-density detections in the galactic plane with relatively low significance and reduce false detections caused by star density fluctuations in the higher latitude region, corrections are carefully made as described in the following procedures. First, we adopted the square root of the star density in each density peak as the weight for this peak, which could help to strengthen the significance of low latitude peaks. Second, the majority of known clusters, either MWSC or DAML02, locate in the interval of the Galactic latitude $-$20$^{\rm o} < b <$ 20$^{\rm o}$. Thus, regions outside this interval are not considered in our work. Third, peaks with low star counts are excluded as well. Finally, there are 1\,921 density peaks detected above the weighted detection threshold, which is 4.5 $\sigma$.

After these over-density peaks were selected, each peak was examined with its $1^{\circ} \times 1^{\circ}$ true color image built from XSTPS-GAC three-band CCD images. For a real cluster, member stars will spatially concentrate in the true color image. Furthermore, the distribution of cluster member stars on the CMD should follow an isochrone-like trend.  Finally, there are 320 known open clusters, either in DAML02 or MWSC, were rediscovered. However, 12 of these known clusters, mainly discovered by \cite{Kronberger2006}, were only cataloged by their celestial coordinates without any physical parameters.  Among other detected unknown peaks, there are 30 peaks classified as high probability new cluster candidates. 

In the following analysis, our new discovered cluster candidates and other 12 known clusters are studied together.  The new clusters will be named by the character $Xuyi$ with a suffix number repesents their discovery order.  While for known clusters, the original names are still used here. All clusters have a IAU-recognized format DSH Jhhmm.m$\pm$ddmm ID assigned.

\subsection{Fundamental Parameters}
In this work, the most concentrated place of star cluster members, i.e. the highest number density point, is considered as the cluster center. We iteratively calculated the mode of star coordinates in a selected 30$\times$30 arcmin$^{2}$ region. The mode defined in the SE$_{\rm XTRACTOR}$ software \citep{Bertin1996} is adopted here, as shown in Equation (\ref{eq:mode}).

\begin{equation}
\label{eq:mode}
\begin{aligned}
\rm Mode = 2.5 \times \rm Median - 1.5 \times \rm Mean \\
\end{aligned} 
\end{equation}

Where Median is the median value of star coordinates in the selected field, and Mean is the mean value of star coordinates.

  After the accurate estimations of the cluster central coordinates were obtained, we created radial density profile (RDP) by calculating star number density in concentric rings. The RDP was then used to check whether a cluster candidate exhibit an obvious excess in the star number compared to the background. The RDP was used to determine the radius of the candidate cluster as well. 

The core radius ($R_{core}$) of each cluster candidate was acquired by fitting its RDP to the empirical King model \citep{King1962}. Then $2\times R_{core}$ was adopted as the cluster visual radius $R$. Outside the visual radius, RDP slope becomes flat and stars within the visual radius are used for cluster parameters analysis. 
However, for faint and diffuse candidates, e.g. $R$~smaller than 1.5$^{'}$, whose fitted $R_{core}$ is so small that not enough stars could be used in the CMD analysis and cluster parameter derivation. While for candidates with large $R_{core}$, there would be many field stars contaminate the analysis. Finally, the stars used for the cluster parameter analysis are selected in the range of 1.5$^{'} < R <$ 4.0$^{'}$.

For each candidate, we selected stars in a square region of 1.0\,deg$^{2}$ around its calculated center. Photometric data from XSTPS-GAC, 2MASS and UCAC4 are used here.  Based on its center coordinate and cluster radius determined above, we built optical CMDs of ($r,\,g-r$), ($i,\,r-i$) and  Infrared CMD of ($J,\,J-H$) for each cluster. The stars within the adopted radius $R$ are used as member candidates and  all stars outside the 2R are used as the control field stars. 

The estimations of age, distance, and reddening for a cluster were done by fitting its CMD to a solar-metallicity isochrone from \cite{Girardi2002}. Mostly, the isochrone fitting was done on the optical CMDs of $(r, g-r)$ and $(i, r-i)$. The comparison were also made on 2MASS $(J, J-H)$ CMD, especially for clusters whose bright stars, such as red giant branch stars, are saturated on XSTPS-GAC. In addition, only for large clusters, we adopted proper motions from UCAC4 to choose relative robust cluster members before isochrone fitting.  The final parameters were determined from the best fitting on the $(i, r-i)$ CMD.  The parameter uncertainties were adopted from derived parameter differences between CMD of $(i,r-i)$ and $(r, g-r)$.

Most known open clusters toward Galactic anti-center are relatively old, and each candidate discovered here is thought to be old enough to show its turn-off phase stars on the CMD. Therefore, by comparing with background distribution, we identified the cluster--like feature, then manually picked the approximate CMD coordinate of turn-off stars as the initial offsets for each isochrone fitting. The adjustment of the theoretical isochrones on CMD along the vertical axis represents the change of cluster visual distance module, meaning the cluster reddening along the horizontal axis.

Even though there are several automatic isochrone fitting methods as mentioned in the introduction, the isochrone fitting in this work was done manually. One reason is that almost all of these methods are fit for well observed clusters, e.g. with accurate and deep photometric measurement for plenty of cluster members, or with well measured proper motion. However, for diffuse or faint clusters, these approaches will produce large error. Furthermore, each of these automatic methods will return a series of suitable parameter sets, which also need one to determine the best set.

The isochrone fitting was carried out in the following steps.  First, isochrones of log($age$) = 7.00, 7.50, 8.00, 8.50, 9.00, 9.50 were picked out as basic isochrone set. Then through the rough fitting to basic isochrones on the CMD, we chose the best one as cluster's reference isochrone, and referred to this corresponding age as $age_{ref}$.  Second, based on the reference, isochrones between $age_{ref} - 0.3$ and $age_{ref} +0.3$  in the step of 0.05 were adopted. In order to perform a more accurate isochrone fitting, the isochrone were shifted along the color and magnitude axes with steps of 0.02 and 0.1 mag, respectively. Typically, after tens of fitting steps, the parameters of a cluster can be properly determined. Then we repeated the same steps on other CMDs independently for comparison, so that we could find the best isochrone fitting and corresponding cluster parameters.  

\begin{table}
\caption{Extinction coefficients used in this papaer, in the unit of $E(B-V)$.}
\label{extins}
\begin{tabular}{c|c|c|c}
\hline
Filter & Value & Color & Value \\ \hline
 $r$   & 2.31  & $g-r$ & 0.99  \\
 $i$   & 1.71  & $r-i$ & 0.60  \\ 
 $j$   & 0.72  & $j-h$ & 0.26  \\ \hline
\end{tabular}
\end{table}

Reddening law from \cite{Yadav2011} was used to derive $E(B-V)$ and $A_V$ for each cluster, 
and the extinction coefficients from \cite{Yuanext2013}, shown in Table\,1, were adopted.   \cite{Yuanext2013} 
carefully performed reddening measurement based on photometric data from GALEX, SDSS, 2MASS and WISE, 
ranging from the far ultraviolet to the mid--infrared. More details could be found in that paper.

\section{Results and Discussions}

In this work, great effort has been spent to inspect the 30 overdensity peaks, also 12 known clusters, and to derive their physical parameters as well. However, the physical parameters of 6 new cluster candidates and 1 known cluster can not be derived properly, owing to the significant error in isochrone fitting. Apart from these, the remaining 24 new stellar overdensities turn out to be new cluster candidates and their parameters are listed in Table 2.  The remaining 11 known clusters could be well studied and their parameters are listed in Table 3. All the well studied clusters are exhibited in Fig\,\ref{fig1} -- \ref{fig10} with their CCD images, RDPs and $(r, g-r)$, $(i, r-i)$ and $(J, J-H)$ CMDs. For each CMD, the stars within the adopted radius $R$ are considered as member candidates and displayed in filled black circles, while the stars outside the $2\times$~R are considered as the control field stars and displayed in black dots. 

\vspace{3ex}
$Obvious~Clusters.$ Teutsch 4, Alessi 50, Juchert 23 are three obvious clusters in this study, shown clear spatial concentration and could be well separated from the background stars. On their CMD, cluster member candidates have shown clear main sequence and giant branch trends.  Parameters of these three clusters are studied for the first time.
In Fig\,\ref{fig1} we illustrated one of the obvious clusters Alessi 50, with clear main sequence and giant branch stars. 
The labeled parameters derived from the $(i, r-i)$ CMD are in good agreement with the parameters derived from the other two CMDs. The parameters and their uncertainties can be found in Table\,3.  Juchert 23 is relatively diffuse compare to other two open clusters, but still could be detected from background.  Alessi 50 is the closest among the three, suffering less dust reddening, while Teutsch 4 and Juchert 23 are located beyond 3.0\,kpc.

The second kind of obvious clusters is relatively compact cluster. These open clusters are dense and could be clearly seen on the CCD images. However, their main sequence stars are heavily mixed with field stars and their bright giant member stars are saturated on CCD. Examples are Xuyi 24 as shown in Fig\,\ref{fig2}, and other clusters like Teutsch 60, Teutsch 20, Teutsch 59b and Xuyi 01. One should pay careful attention to these problems when identifying cluster features on their optical CMDs. 
Fortunately, there are some clear giant stars on their infrared CMDs, which could help to constrain the accuracy of parameters. 
The parameters derived on the $(J,J-H)$ CMD are used as reference to determine their labeled parameters on $(i, r-i)$ CMD.
The determined parameters show that all the five clusters are old open clusters, around 1\,Gyr, and their distances are ranging from 1.5\,kpc to 3.2\,kpc.

\vspace{3ex}
$Obscured~Clusters.$ Three faint clusters, Teutsch 58, Xuyi 21 and Xuyi 23, were discovered in this work. But they are still dense enough to be evident on their images. Teutsch 58 is very red on its CCD image.  Xuyi 21 is likely to be hidden by foreground stars. Xuyi 23 is very close to known cluster vdBergh\_92, which is a large and bright cluster, as shown in the left panel of Fig\,\ref{fig5}.  Red giant stars are detected in all three clusters, but their main sequence stars are contaminated by faint background stars. After analysis were carried out carefully, their ages were derived to be about 1\,Gyr.

\vspace{3ex}
$Bright~Clusters.$ Three bright clusters, Xuyi 10, Xuyi 11, Xuyi 20, are studied here.  They were thought to be clusters of spikes at first, because their members are too bright. Fortunately they were picked out by visual inspection later. One bright cluster, Xuyi 20, is shown in Fig\,\ref{fig6}. Their members are saturated on the XSTPS-GAC image, so 2MASS data were used to help deriving their parameters. In addition, their ages are relatively young.  Xuyi 11 turns out to be a very young cluster, with the age of about 100\,Myr. Like clusters in the blue main sequence group, the main sequence of Xuyi 20 exhibits a very blue feature. Proper motions from UCAC4 were used to identify member stars and sharp the cluster features on CMDs.

\begin{table*}
\centering
\begin{minipage}{140mm}
\caption{Parameters of new clusters.}
\label{clusters_new}
\begin{tabular}{c|c|c|c|c|c|c|c}
\hline
\hline
  DSH ID          & Name        & RA J2000 & Dec J2000  &R$_{core}$& log Age           & $E(B-V)$      & Distance          \\ 
                  &             & hh mm ss &  dd mm ss  &arcmin& $log\,yr$       &  mag          & kpc               \\ \hline
DSH J0314.8+5912  & Xuyi 01     & 03 14 49 &  +59 12 56 & 1.98 & 8.85$\pm$0.05 & 0.82$\pm$0.44 &  2.09$\pm$0.30  \\
DSH J0352.1+5230  & Xuyi 02     & 03 52 04 &  +52 30 07 & 1.52 & 8.75$\pm$0.30 & 0.98$\pm$0.40 &  1.72$\pm$0.30  \\
DSH J0353.2+5426  & Xuyi 03     & 03 53 11 &  +54 26 38 & 0.84 & 8.70$\pm$0.40 & 1.09$\pm$0.02 &  2.64$\pm$0.60  \\
DSH J0425.3+4901  & Xuyi 04     & 04 25 17 &  +49 01 04 & 0.99 & 8.75$\pm$0.30 & 1.41$\pm$0.30 &  2.93$\pm$0.26  \\
DSH J0439.3+4811  & Xuyi 05     & 04 39 20 &  +48 11 57 & 1.17 & 8.95$\pm$0.30 & 1.36$\pm$0.18 &  3.80$\pm$0.95  \\
DSH J0451.4+3859  & Xuyi 06     & 04 51 23 &  +38 59 29 & 0.76 & 9.45$\pm$0.30 & 0.90$\pm$0.38 &  2.12$\pm$0.10  \\
DSH J0456.5+5017  & Xuyi 07     & 04 56 27 &  +50 17 41 & 0.71 & 9.00$\pm$0.20 & 0.91$\pm$0.22 &  4.44$\pm$0.68  \\
DSH J0518.7+4510  & Xuyi 08     & 05 18 41 &  +45 10 04 & 1.84 & 9.20$\pm$0.30 & 0.52$\pm$0.24 &  3.44$\pm$0.11  \\
DSH J0608.3+2804  & Xuyi 09     & 06 08 18 &  +28 04 13 & 3.09 & 8.55$\pm$0.15 & 0.37$\pm$0.08 &  5.78$\pm$1.92  \\
DSH J0614.9+1645  & Xuyi 10     & 06 14 57 &  +16 45 06 & 7.78 & 8.40$\pm$0.10 & 0.93$\pm$0.05 &  3.46$\pm$0.79  \\
DSH J0625.7+1109  & Xuyi 11     & 06 25 41 &  +11 09 56 & 8.93 & 7.85$\pm$0.15 & 0.39$\pm$0.02 &  1.46$\pm$0.11  \\
DSH J0628.3+1135  & Xuyi 12     & 06 28 16 &  +11 35 13 & 1.42 & 8.80$\pm$0.20 & 0.78$\pm$0.26 &  4.53$\pm$0.28  \\
DSH J0628.4+1225  & Xuyi 13     & 06 28 23 &  +12 25 54 & 1.62 & 8.70$\pm$0.10 & 0.62$\pm$0.19 &  3.31$\pm$0.57  \\
DSH J0632.1+1605  & Xuyi 14     & 06 32 04 &  +16 05 23 & 1.24 & 8.85$\pm$0.20 & 0.57$\pm$0.30 &  5.57$\pm$0.52  \\
DSH J0634.6--0851 & Xuyi 15     & 06 34 37 & --08 51 59 & 1.69 & 8.85$\pm$0.15 & 0.58$\pm$0.37 &  7.59$\pm$0.57  \\
DSH J0642.4--0130 & Xuyi 16     & 06 42 23 & --01 30 47 & 0.95 & 8.75$\pm$0.10 & 0.45$\pm$0.12 &  5.46$\pm$0.22  \\
DSH J0643.2--0451 & Xuyi 17     & 06 43 13 & --04 51 46 & 1.73 & 8.55$\pm$0.05 & 0.48$\pm$0.12 &  9.16$\pm$1.23  \\
DSH J0643.9+0116  & Xuyi 18     & 06 43 52 &  +01 16 55 & 1.61 & 8.80$\pm$0.05 & 0.50$\pm$0.02 &  1.88$\pm$0.13  \\
DSH J0644.9--0926 & Xuyi 19     & 06 44 51 & --09 26 02 & 0.98 & 8.75$\pm$0.15 & 0.56$\pm$0.03 &  5.99$\pm$1.09  \\
DSH J0649.5+1202  & Xuyi 20     & 06 49 28 &  +12 02 46 & 1.46 & 8.30$\pm$0.15 & 0.35$\pm$0.06 &  3.99$\pm$0.86  \\
DSH J0651.9--1127 & Xuyi 21     & 06 51 55 & --11 27 42 & 0.66 & 8.90$\pm$0.20 & 1.37$\pm$0.42 &  5.58$\pm$0.06  \\
DSH J0702.8--0204 & Xuyi 22     & 07 02 49 & --02 04 25 & 2.68 & 8.55$\pm$0.15 & 0.26$\pm$0.03 &  2.86$\pm$0.16  \\
DSH J0704.6--1115 & Xuyi 23     & 07 04 33 & --11 15 33 & 2.75 & 8.85$\pm$0.15 & 0.90$\pm$0.14 &  7.44$\pm$0.94  \\
DSH J0705.4--0903 & Xuyi 24     & 07 05 24 & --09 03 56 & 1.69 & 8.70$\pm$0.15 & 0.52$\pm$0.12 &  2.26$\pm$0.47  \\
\hline
\end{tabular}
\end{minipage}
\end{table*}

\begin{table*}
\centering
\begin{minipage}{140mm}
\caption{Parameters of 11 known clusters.}
\label{clusters_kk}
\begin{tabular}{c|c|c|c|c|c|c|c}
\hline
\hline
  DSH ID          & Name        & RA J2000 & Dec J2000  &R$_{core}$& log Age           & $E(B-V)$      & Distance         \\ 
                  &             & hh mm ss &  dd mm ss  &arcmin& $log\,yr $       &  mag          & kpc              \\ \hline
DSH J0438.3+4317  & Teutsch 4   & 04 38 16 &  +43 17 30 & 1.26 & 8.90$\pm$0.10 & 0.72$\pm$0.01 &  3.88$\pm$0.80 \\
DSH J0507.7+1734  & Juchert 23  & 05 07 40 &  +17 34 34 & 3.69 & 8.45$\pm$0.05 & 0.74$\pm$0.07 &  3.46$\pm$0.28 \\
DSH J0548.0+2530  & Teutsch 57  & 05 48 00 &  +25 30 03 & 0.89 & 8.60$\pm$0.10 & 0.72$\pm$0.15 &  6.59$\pm$0.17 \\
DSH J0602.2+2607  & Teutsch 92  & 06 02 15 &  +26 07 29 & 1.17 & 8.10$\pm$0.10 & 0.50$\pm$0.04 &  3.27$\pm$0.60 \\
DSH J0606.6+1557  & Teutsch 58  & 06 06 38 &  +15 57 19 & 0.83 & 9.00$\pm$0.15 & 1.59$\pm$0.36 &  4.11$\pm$1.27 \\
DSH J0622.1+2104  & Alessi 50   & 06 22 05 &  +21 04 31 & 1.07 & 9.10$\pm$0.25 & 0.57$\pm$0.25 &  2.55$\pm$0.20 \\
DSH J0625.8+1954  & Alessi 58   & 06 25 47 &  +19 54 19 & 1.55 & 8.20$\pm$0.10 & 0.50$\pm$0.02 &  3.82$\pm$0.48 \\
DSH J0628.8+1456  & Teutsch 20  & 06 28 48 &  +14 56 00 & 1.20 & 8.80$\pm$0.10 & 0.58$\pm$0.19 &  2.54$\pm$0.05 \\
DSH J0643.8--0052 & Teutsch 59b & 06 43 49 & --00 52 19 & 3.23 & 8.75$\pm$0.15 & 0.60$\pm$0.16 &  3.21$\pm$0.23 \\
DSH J0651.4--0148 & Teutsch 60  & 06 51 22 & --01 49 00 & 1.37 & 9.15$\pm$0.15 & 0.36$\pm$0.02 &  1.55$\pm$0.38 \\
DSH J0715.1--0653 & Patchick 79 & 07 15 09 & --06 53 23 & 2.50 & 8.50$\pm$0.10 & 0.31$\pm$0.04 &  4.59$\pm$0.22 \\

\hline
\end{tabular}
\end{minipage}
\end{table*}

\vspace{3ex}
$Inconspicuous~Clusters.$ 
The clusters Xuyi 02--08,12,14 are inconspicuous clusters, identified by some weak cluster-like features, but heavily contaminated by background stars. Two of them are shown in Fig\,\ref{fig9} and Fig\,\ref{fig10}. The clusters Xuyi 03, Xuyi 05, Xuyi 06, Xuyi 07, Xuyi 08 and Xuyi 14 are identified by some clear giant branch stars.
The clusters Xuyi 02, Xuyi 04, Xuyi 12 and Xuyi 14 are diffuse clusters. The cluster Xuyi 14
are identified by some turn-off phase stars, while clusters 22-24
are identified by their narrow main sequence. Parameters for these clusters are 
carefully derived, and two times of parameter differences between $(i, r-i)$~and $(r, g-r)$ CMD are adopted as their uncertainties.

\section{Summary}
Based on XSTPS-GAC photometry data, we performed new star cluster searching on the stellar density maps. We added a weight factor to the density peak selection method, which was proven to be effective in removing false cluster candidates caused by star density fluctuations. 24 new open cluster candidates was discovered in the region of Galactic anti--center.  We manually derived parameters for these new clusters, and also derived parameters for another 11 already known clusters found by \cite{Kronberger2006}, who only provided cluster coordinates. 

 \begin{figure}
 \begin{center}
 \includegraphics[angle=90,width=90mm]{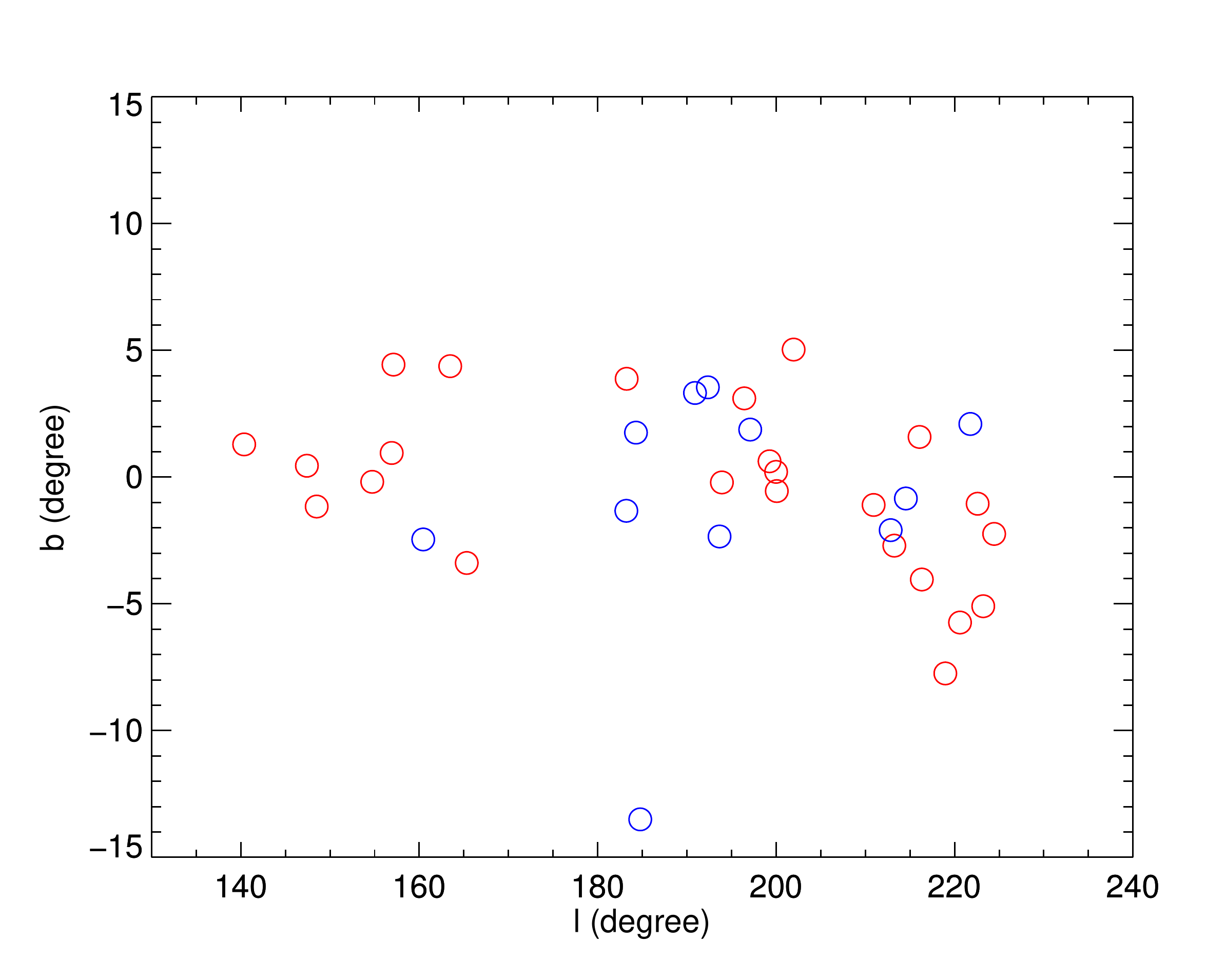}
 \caption{The spatial distribution of all analyzed open cluster candidates in Galactic space. Red circles represent new open cluster candidates discovered in this work, while blue circles are those 11 known open clusters.} 
 \label{fig00}
 \end{center}
 \end{figure}

The obtained cluster parameters may be affected by our subjectivity, which may introduce some uncertainties. Especially for clusters without clear red giant branch or main sequence, like clusters Xuyi 09, Teutsch 58 and Xuyi 19. We took very careful inspection for each cluster on their CMDs, to ensure their member stars could be well fitted, and checked them by using the 2MASS data as well. Three obscured clusters were discovered and studied, indicating that deep infrared sky survey like $VVV$ are needed to find more clusters in the distant Galactic outer disk. And three bright clusters were found as well, Xuyi 10, Xuyi 11 and Xuyi 20, suggesting nearby clusters still could be found by careful searches. Furthermore, more small clusters will be found if we decrease the Gaussian width of the Koposov kernel used to find star density peaks, but it will be difficult to derive cluster parameters for these clusters, since their detected member stars will be just a few. 
On the other hand, almost all the studied clusters in this work could be spectroscopically observed by LAMOST. Then, all newly discovered clusters and known clusters studied in this work will be important probes to study the Galactic dynamics and evolution.

\begin{acknowledgements}
\vspace{2ex}
This work is supported by the National Natural Science Foundation of China
(NSFC) grant \#11473001, \#11233004, \#11078006, \#11633009 and \#11273067, and by the
Minor Planet Foundation of Purple Mountain Observatory. 
This work is also supported by National Key Basic Research Program of China 2014CB845700.  
This work was supported by the China Postdoctoral Science Foundation Grant No.2017M610695.

The LAMOST FELLOWSHIP is supported by Special Funding for Advanced Users, budgeted and administrated by Center for Astronomical Mega-Science, Chinese Academy of Sciences (CAMS).

\end{acknowledgements}

%%%%%%%%%%%%%%%%%%%%%%%%%%%%%%%%%%%%%%%%%%%%%%%%%%%%%%%%%%%%%%%%%%%%%%%%%%%%%%%%%
\bibliographystyle{raa}
\bibliography{RAA-2017-0235}

%======================================================================================================================

 \begin{figure}
 \begin{center}
 \includegraphics[width=160mm]{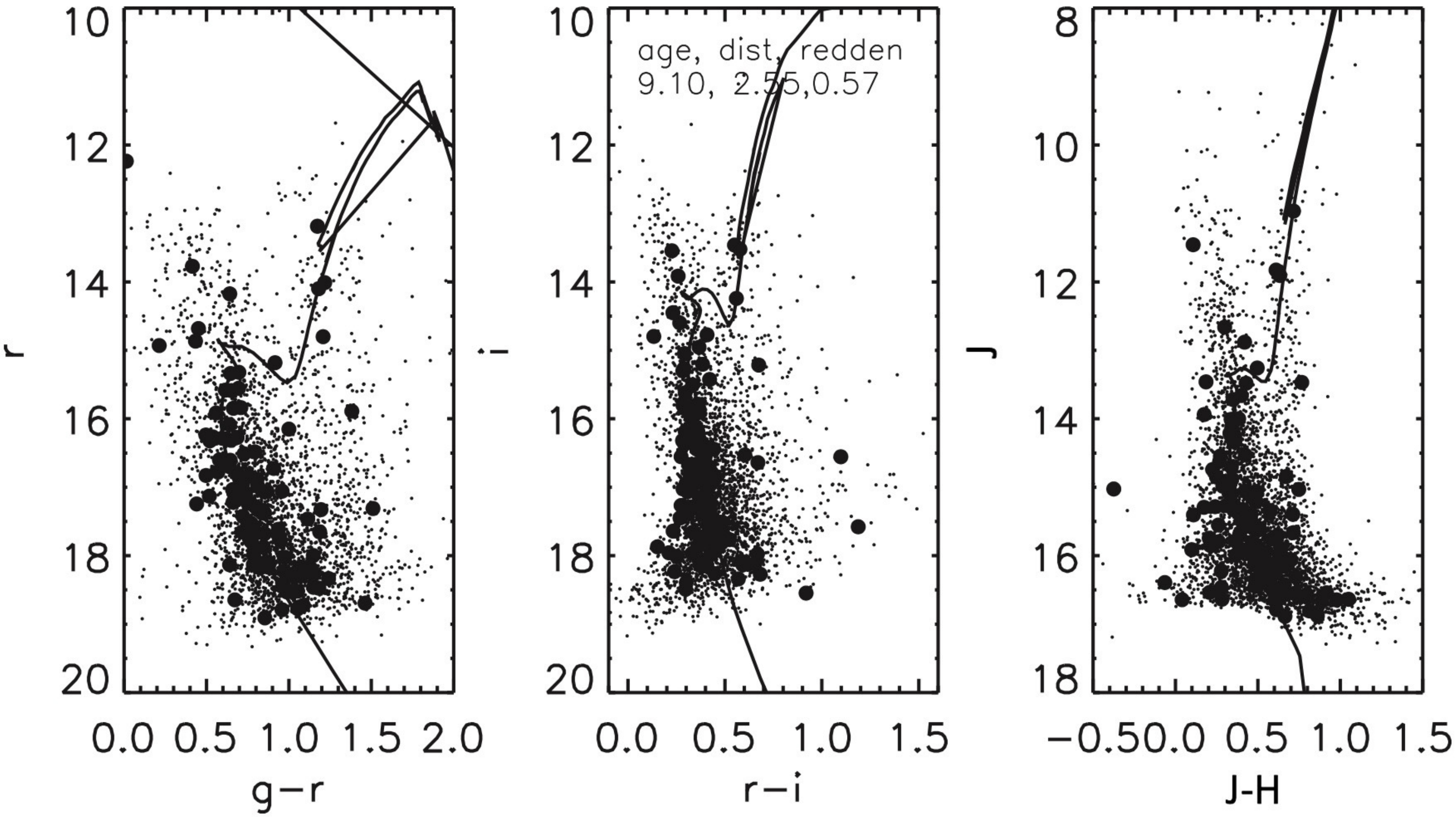}
 \includegraphics[width=60mm]{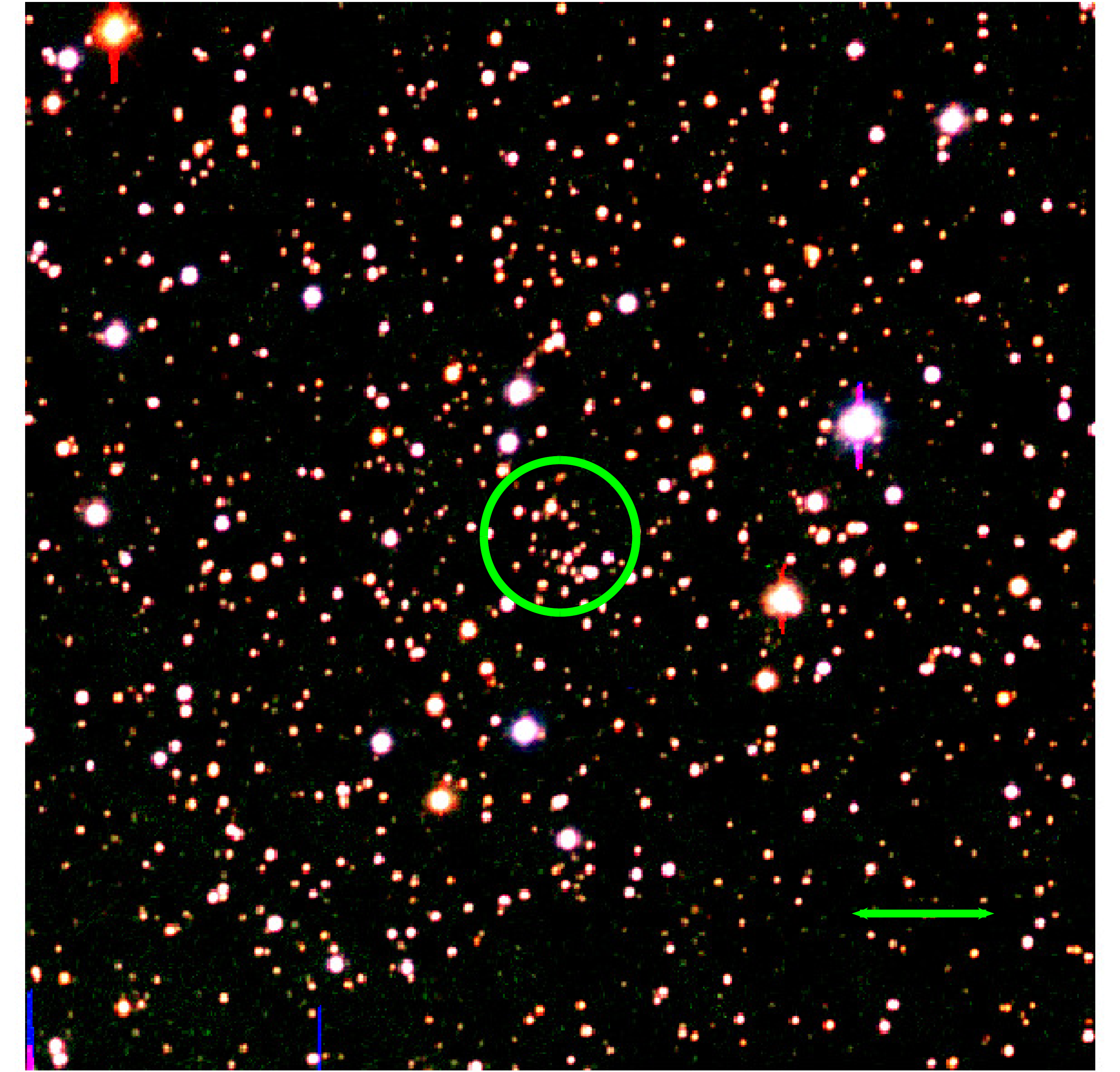}
 \includegraphics[width=80mm]{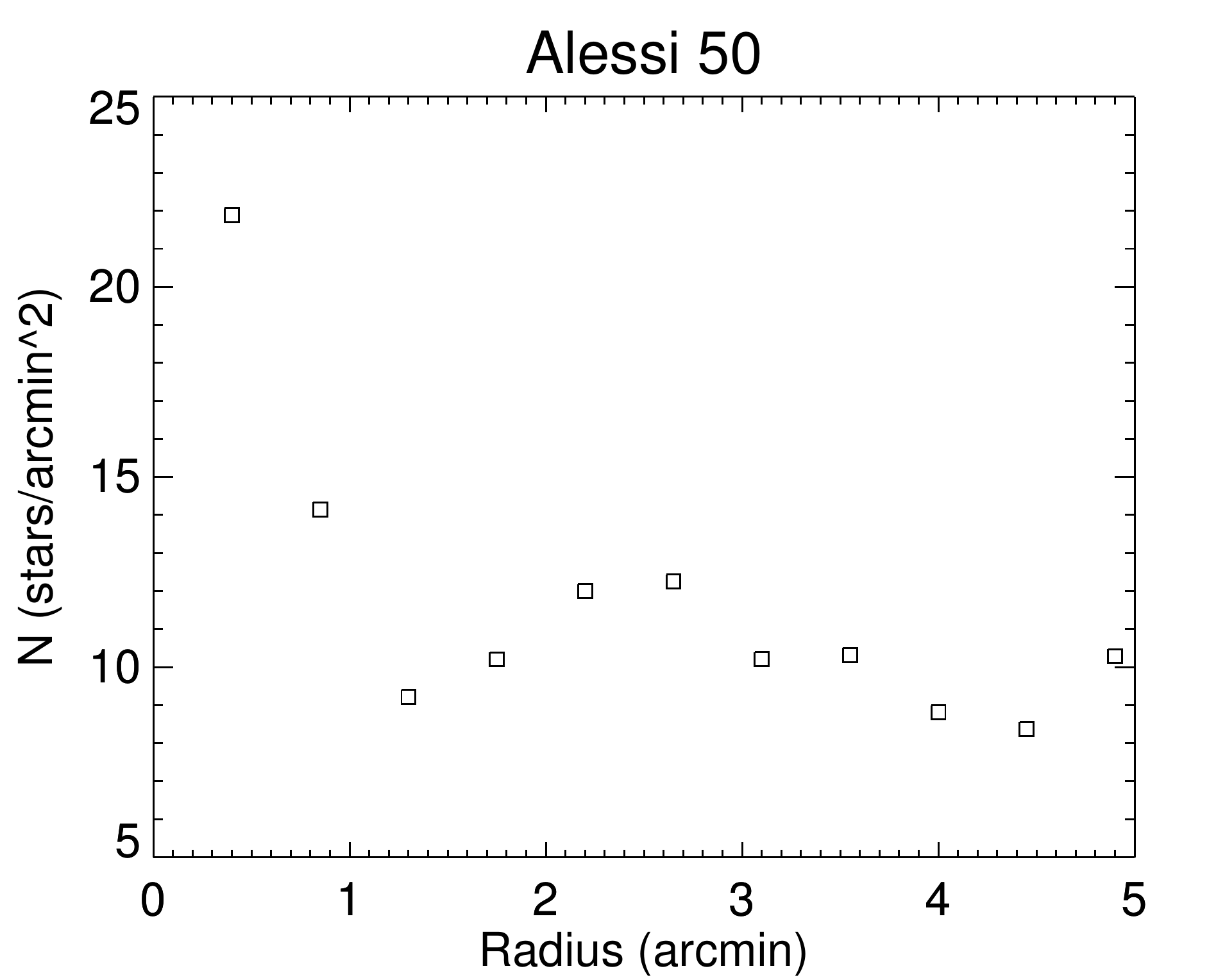}
 \caption{Color--magnitude diagrams, true color image and radial density profile(RDP) for cluster Alessi 50.
The top three panels are CMDs, while bottom left is true color CCD image, and RDP is at the bottom right.
The three CMDs are the $(r, g-r)$, $(i, r-i)$ and 2MASS $(J, J-H)$ diagrams.
The black filled circles on each CMD are the stars within cluster visual radius which 
calculated by radial profile fitting, as described in Sec.\,3.2, and the background dots are 
the conparsion field stars.  
Each true color image was combined by g, r, i band 15$^{'}\times$15$^{'}$ XSTPS-GAC image centering on the cluster center, and the cluster 
core radius $R_{core}$ is labeled by a circle. The line at lower right is a 2$^{'}$ ruler.
For each cluster, the determined log($age$) is labeled on its  $(i, r-i)$ CMD.} 
 \label{fig1}
 \end{center}
 \end{figure}

 \begin{figure*}
 \begin{center}
 \includegraphics[width=170mm]{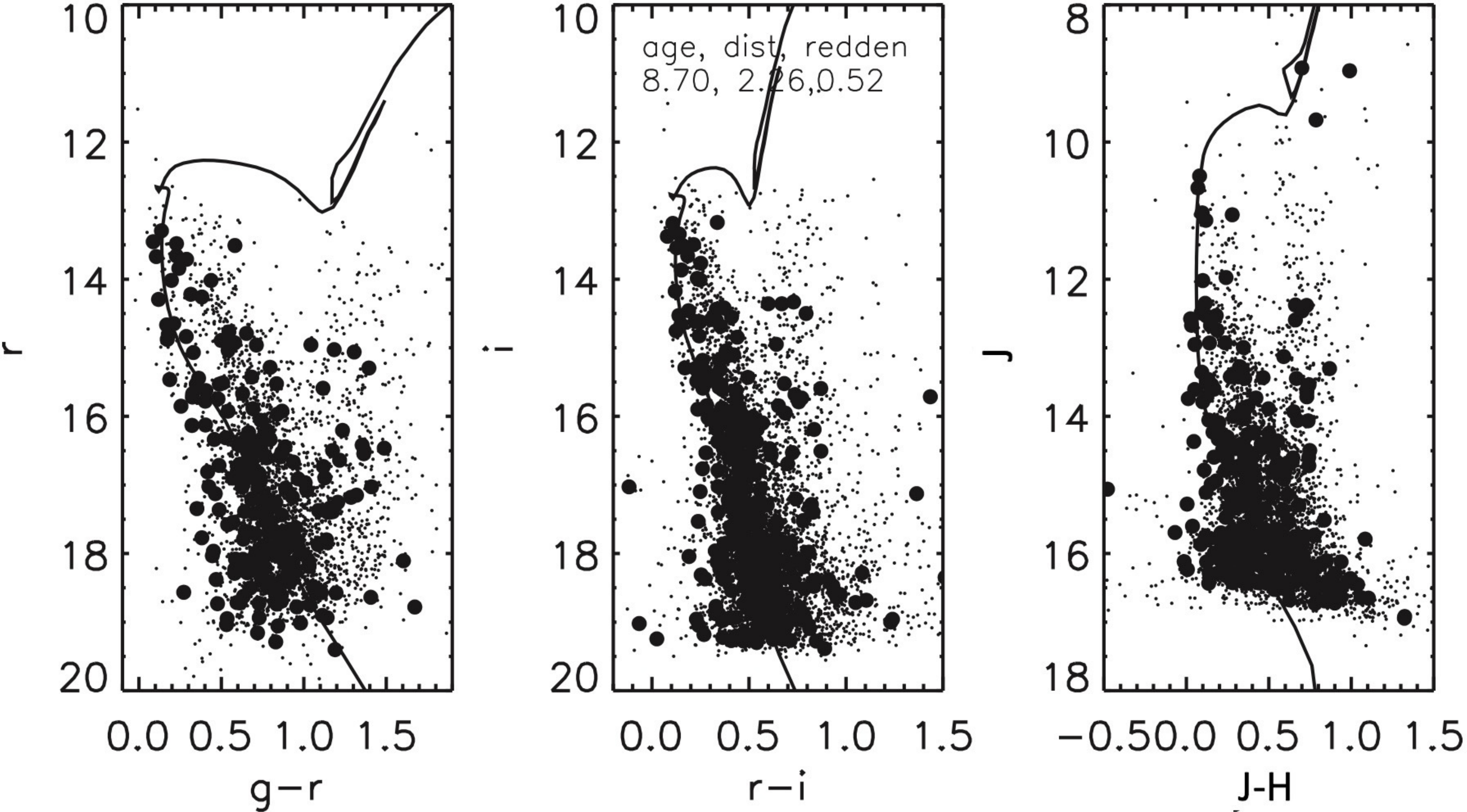}
 \includegraphics[width=60mm]{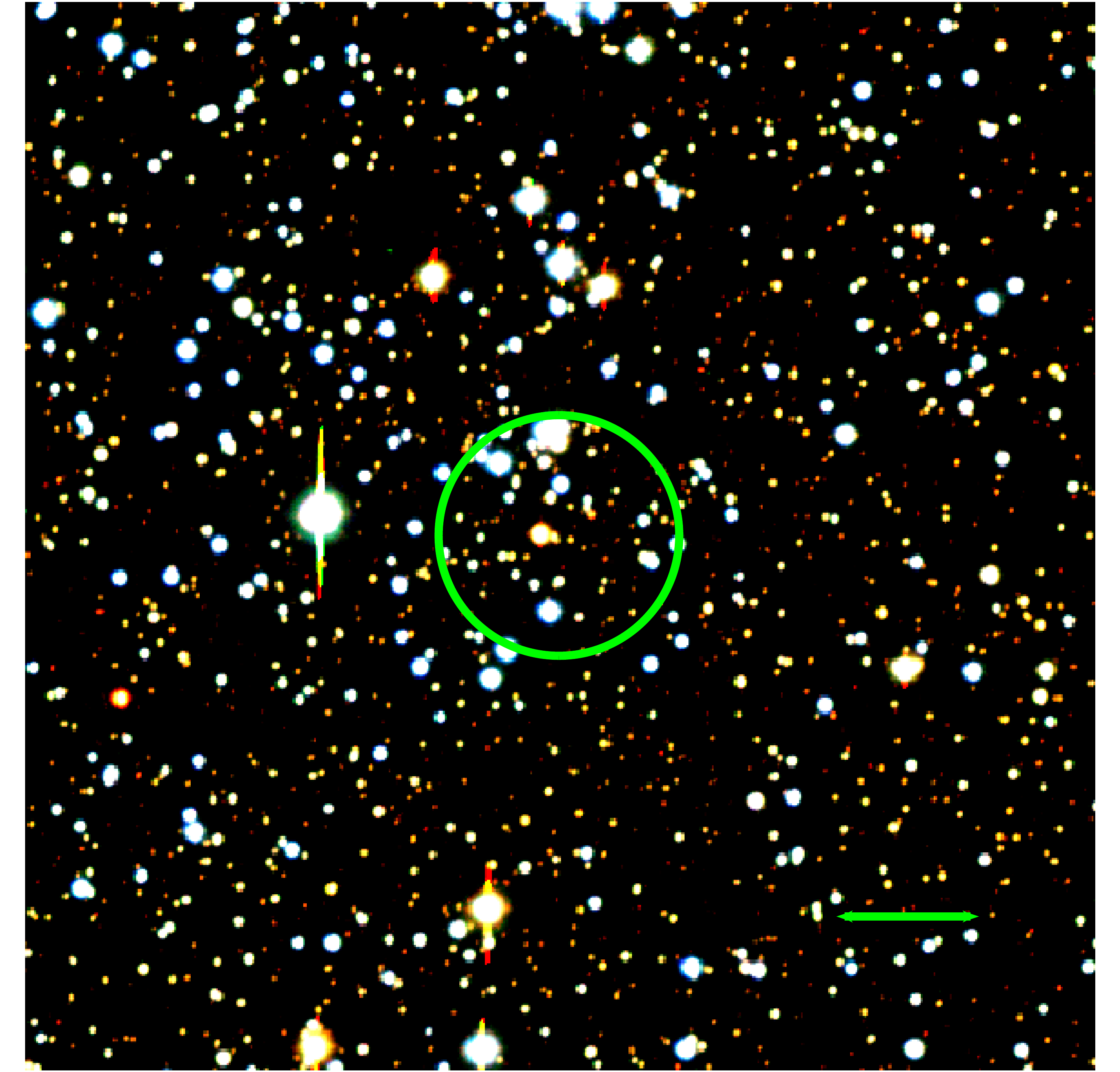}
 \includegraphics[width=80mm]{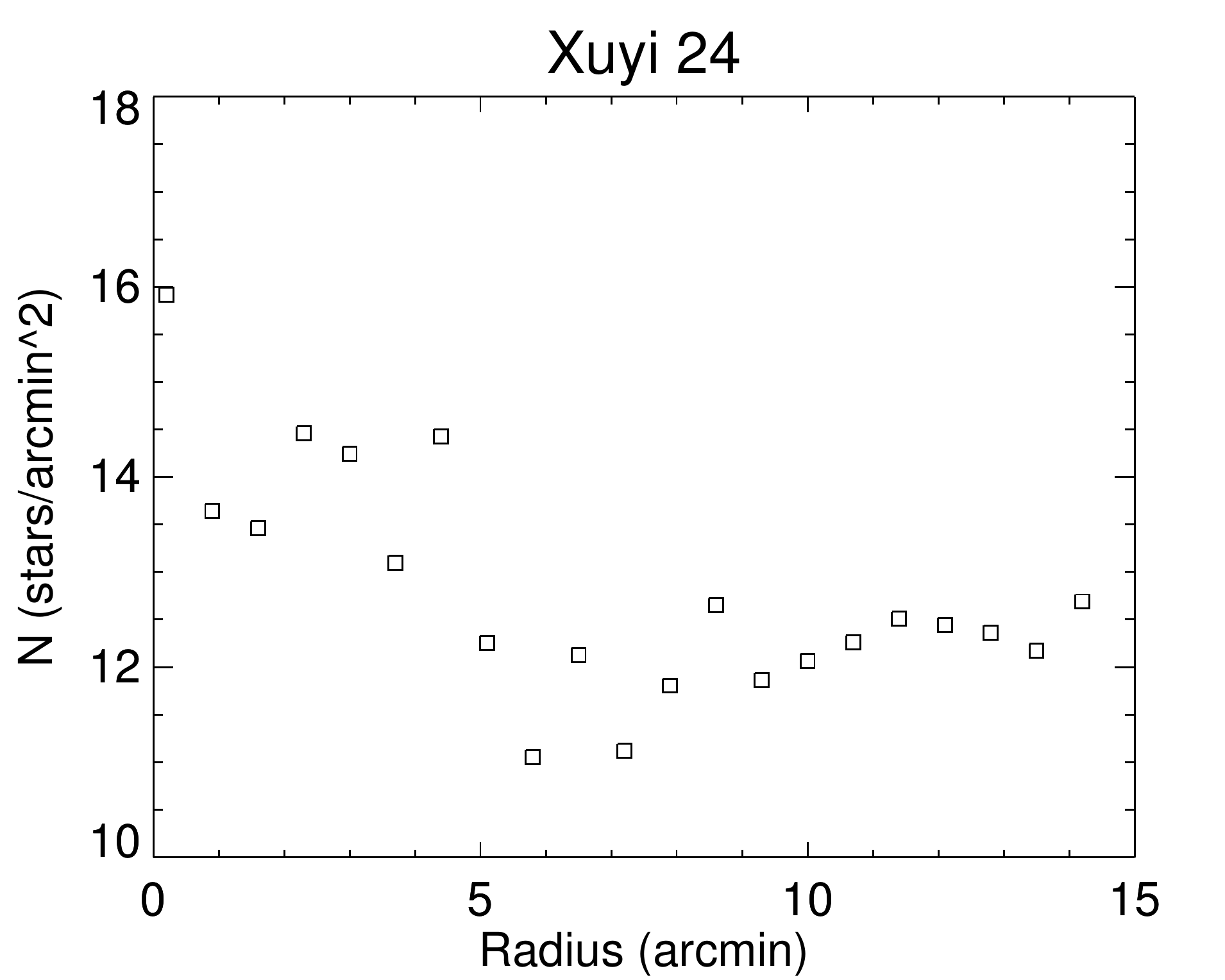}
 \caption{Same as Fig\,\ref{fig1} but for cluster Xuyi 24. It is also an example of
 second obvious clusters that are relatively compact clusters.} 
 \label{fig2}
 \end{center}
 \end{figure*}
 
  \begin{figure*}
 \begin{center}
 \includegraphics[width=170mm]{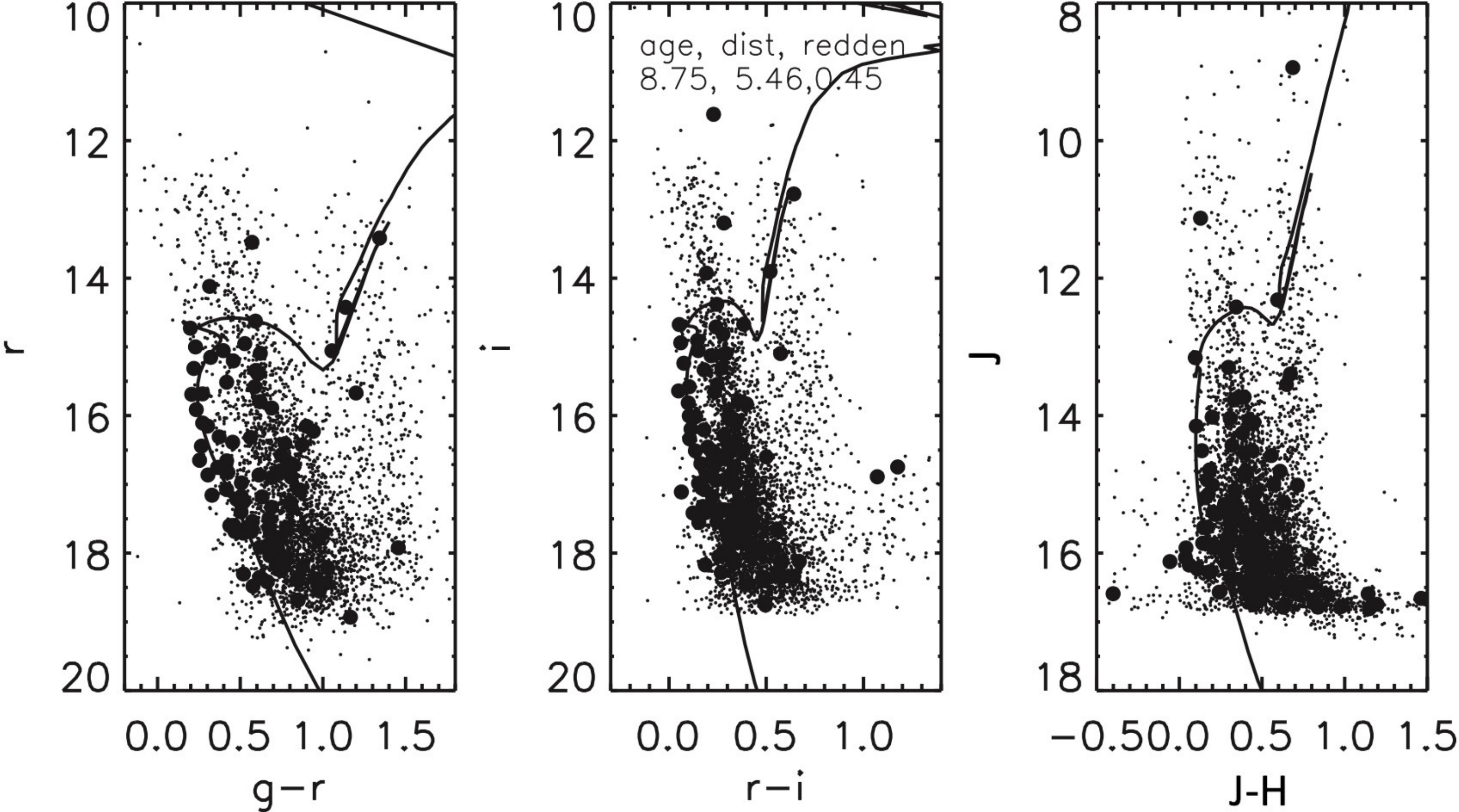} 
 \includegraphics[width=60mm]{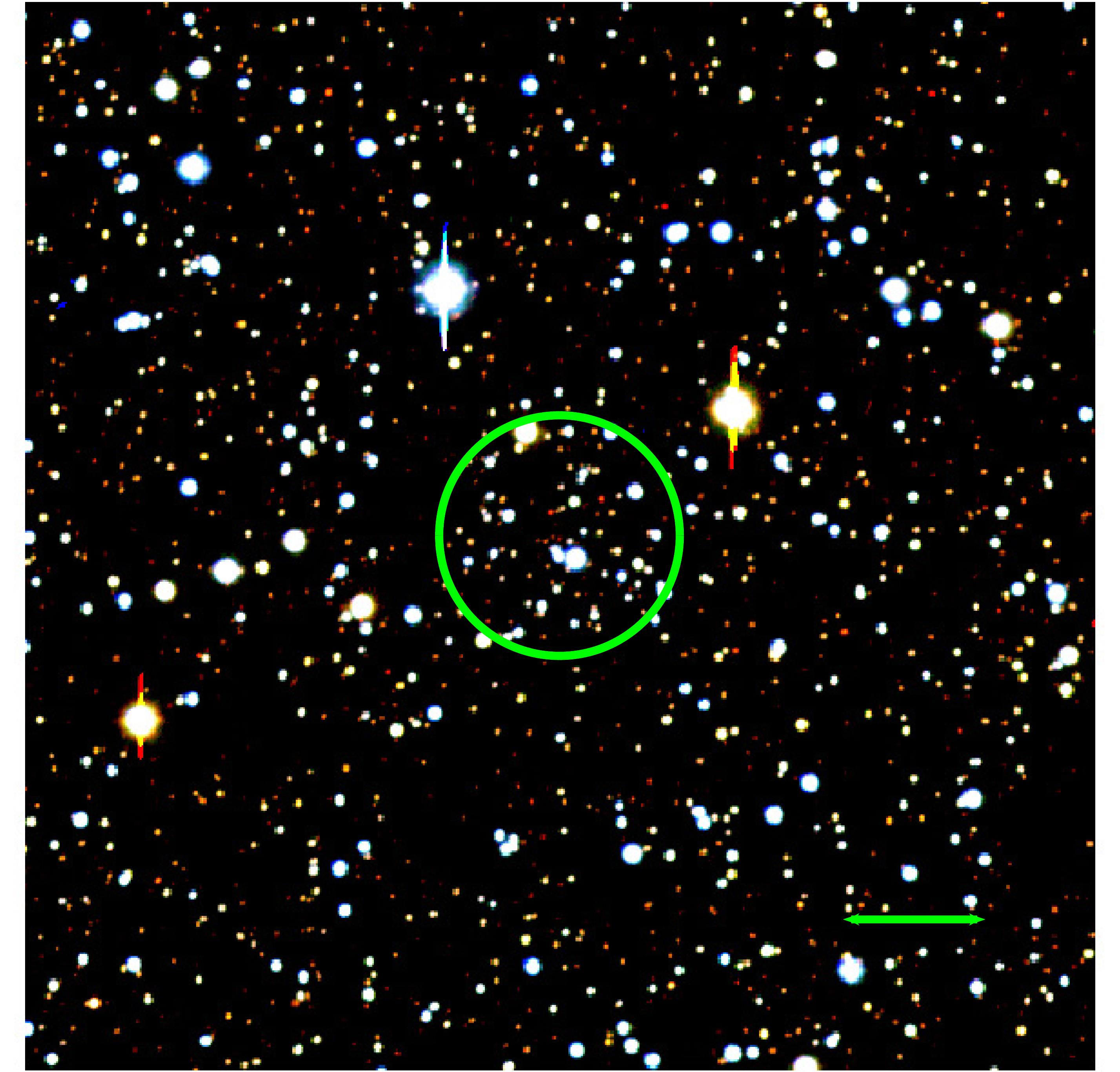}
 \includegraphics[width=80mm]{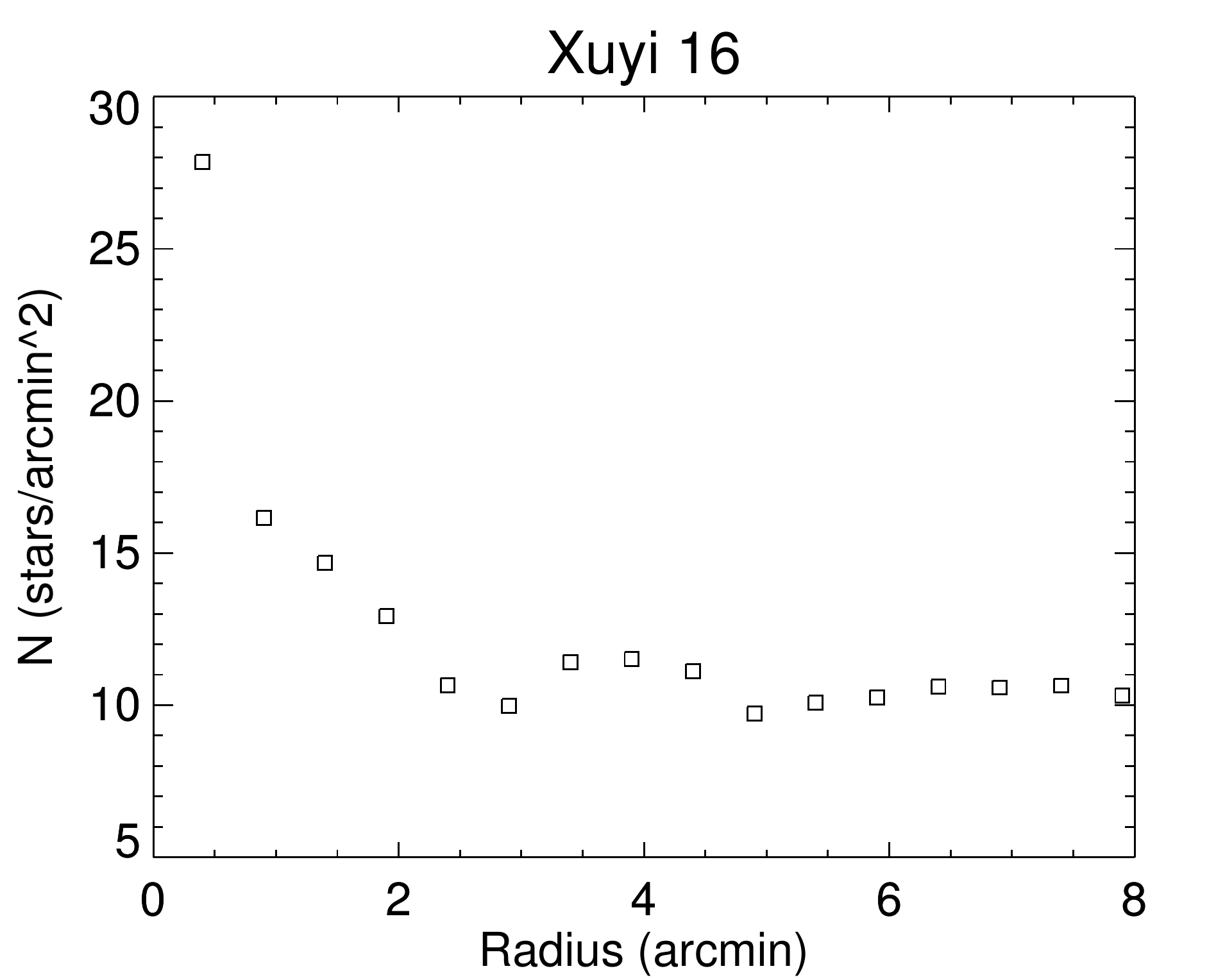}
 \caption{Same as Fig\,\ref{fig1} but for cluster Xuyi 16. It is one example of
 third obvious clusters that show a clear blue main sequence.} 
 \label{fig3}
 \end{center}
 \end{figure*}

 \begin{figure*}
 \begin{center}
  \includegraphics[width=170mm]{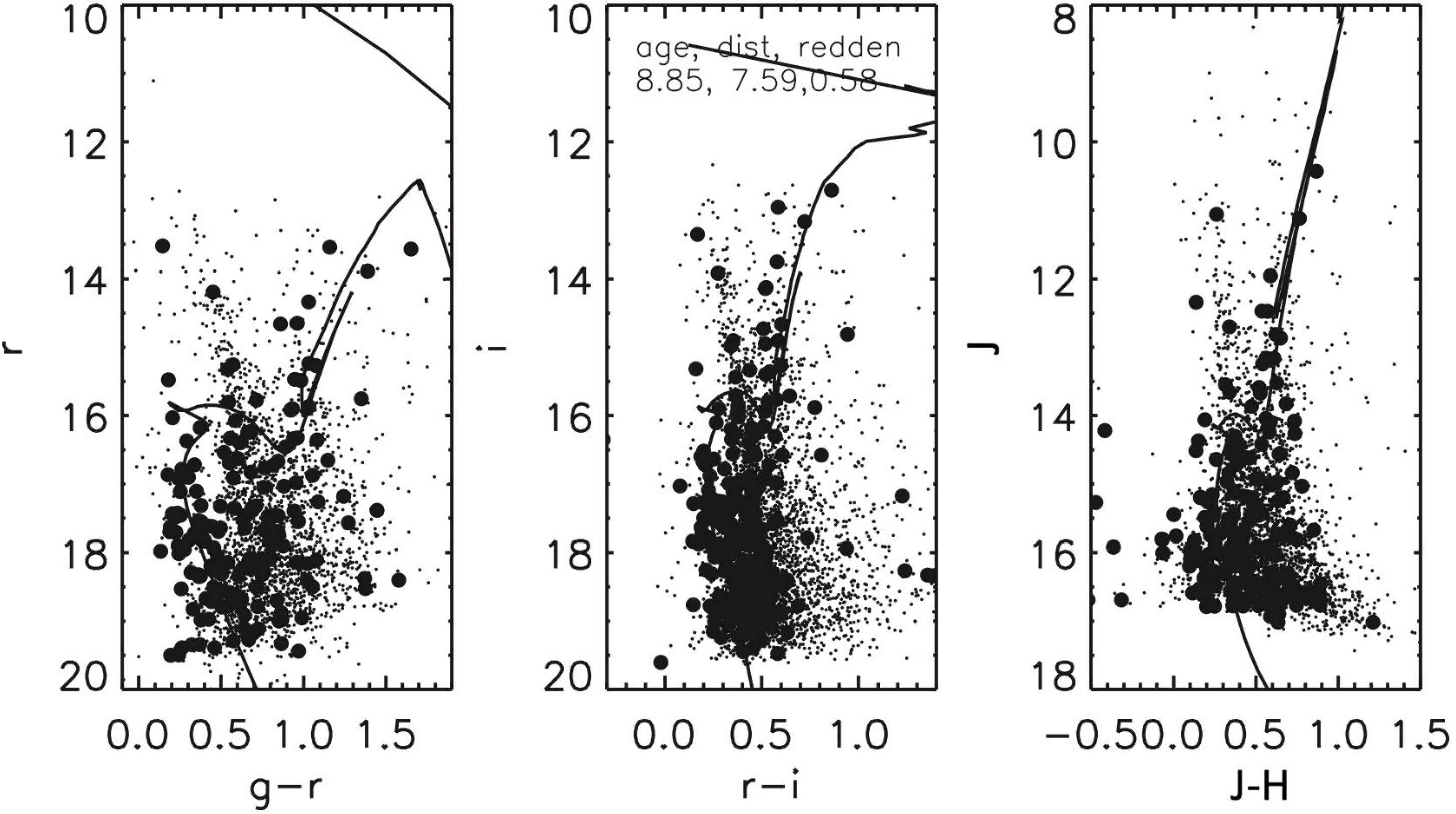}
 \includegraphics[width=60mm]{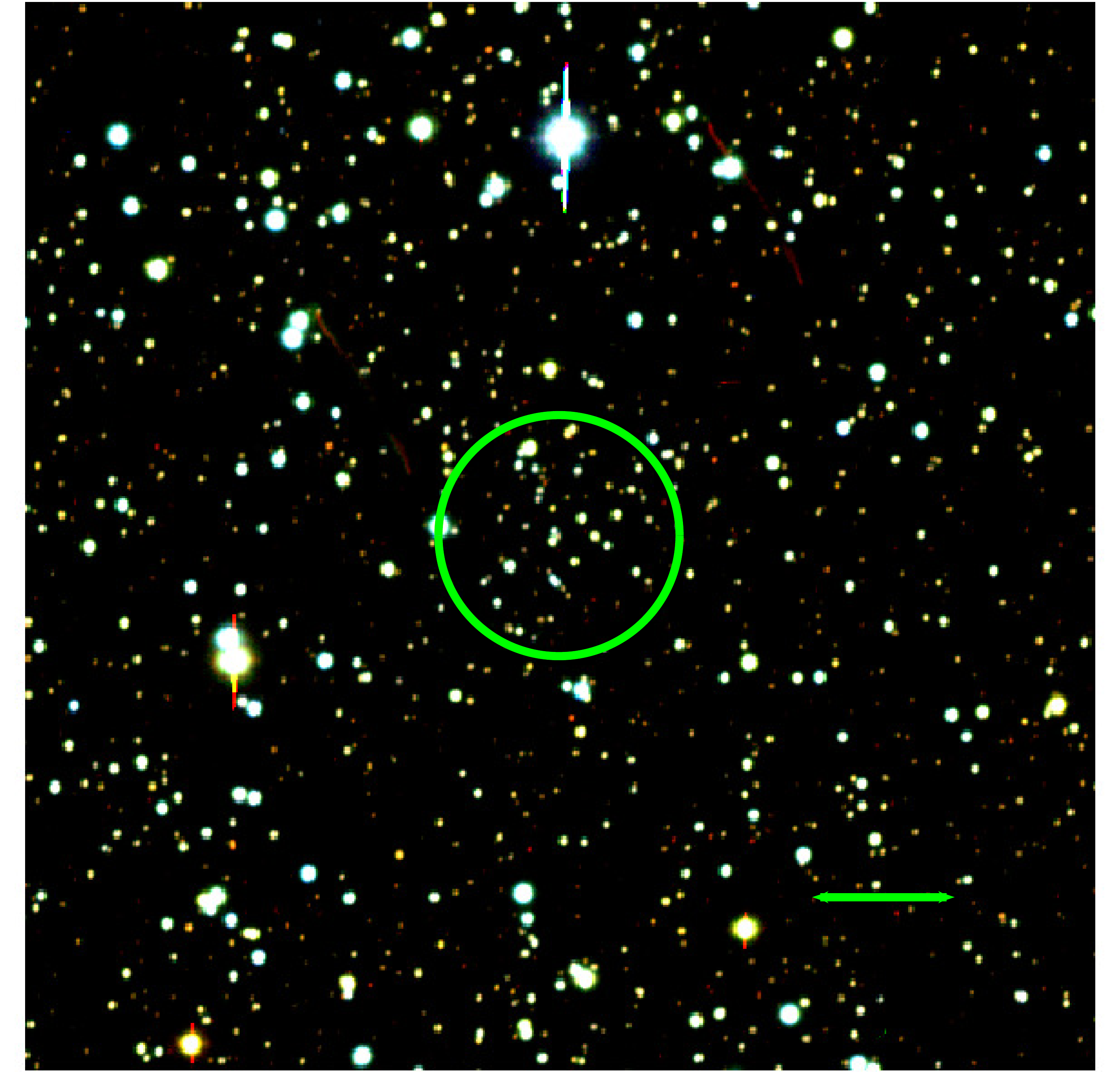}
  \includegraphics[width=80mm]{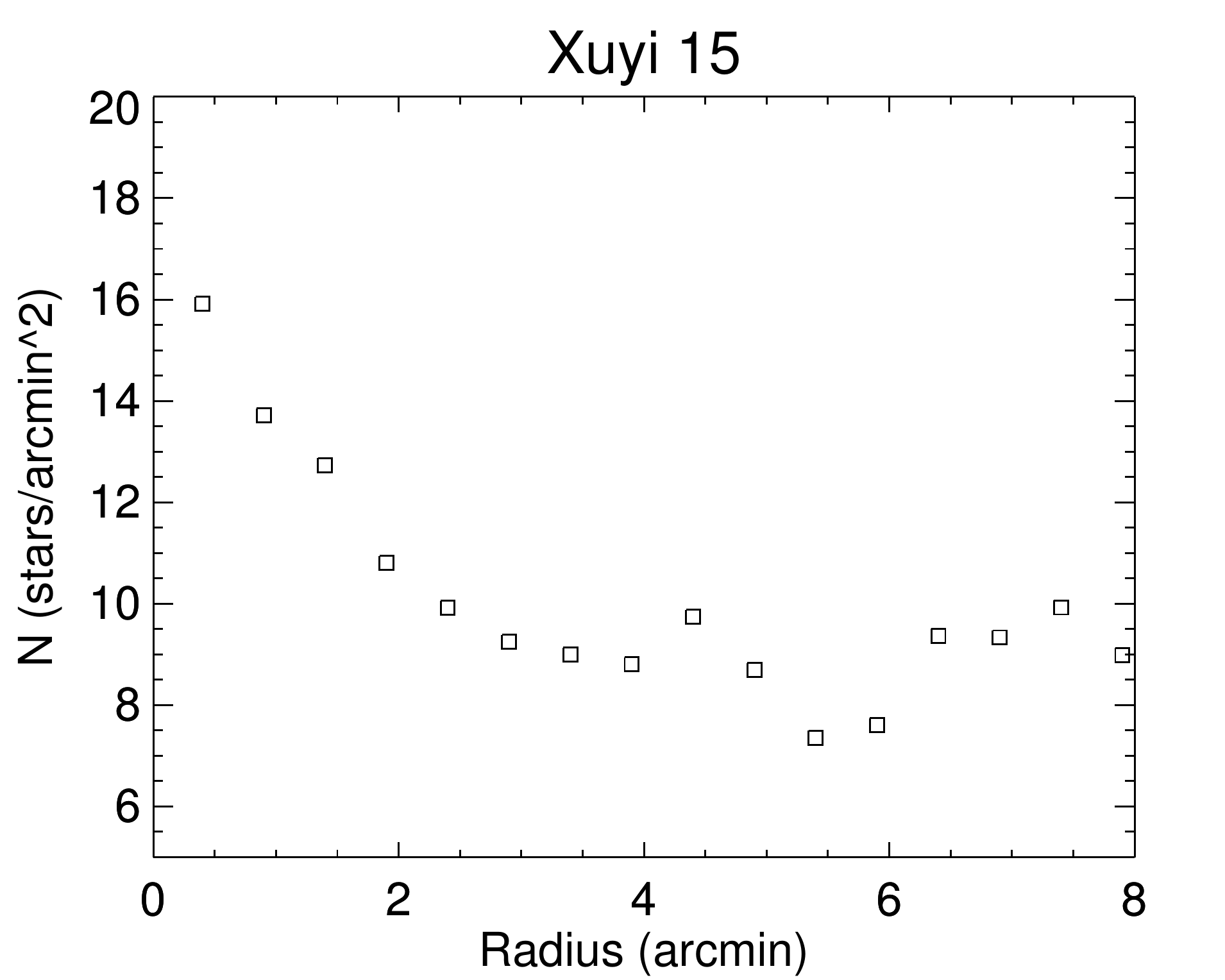}

 \caption{Same as Fig\,\ref{fig1} but for cluster Xuyi 15. This is another example of
 third obvious clusters that show a clear blue main sequence.} 
 \label{fig4}
 \end{center}
 \end{figure*}

 \begin{figure*}
 \begin{center}
 \includegraphics[width=170mm]{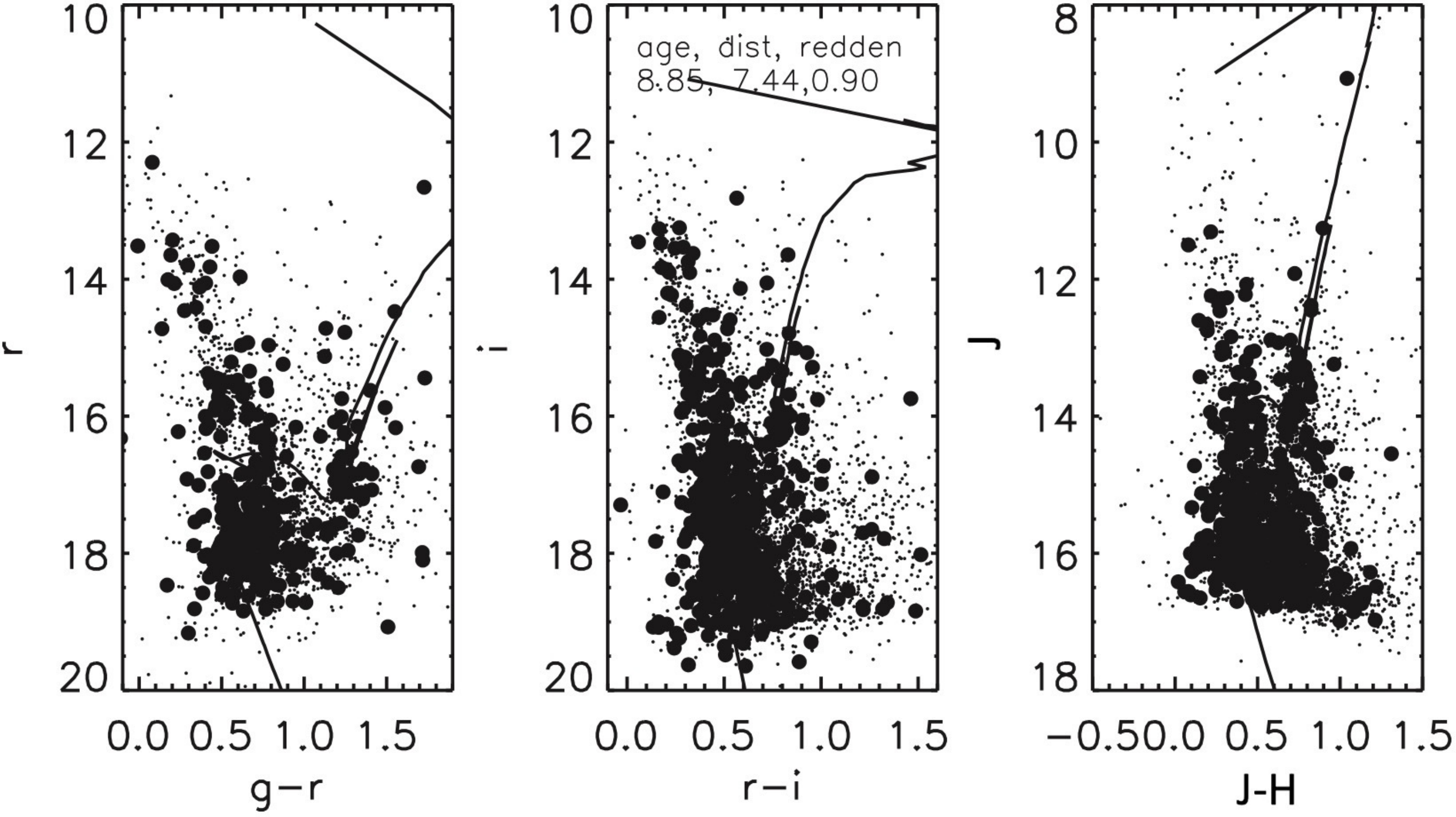}
 \includegraphics[width=60mm]{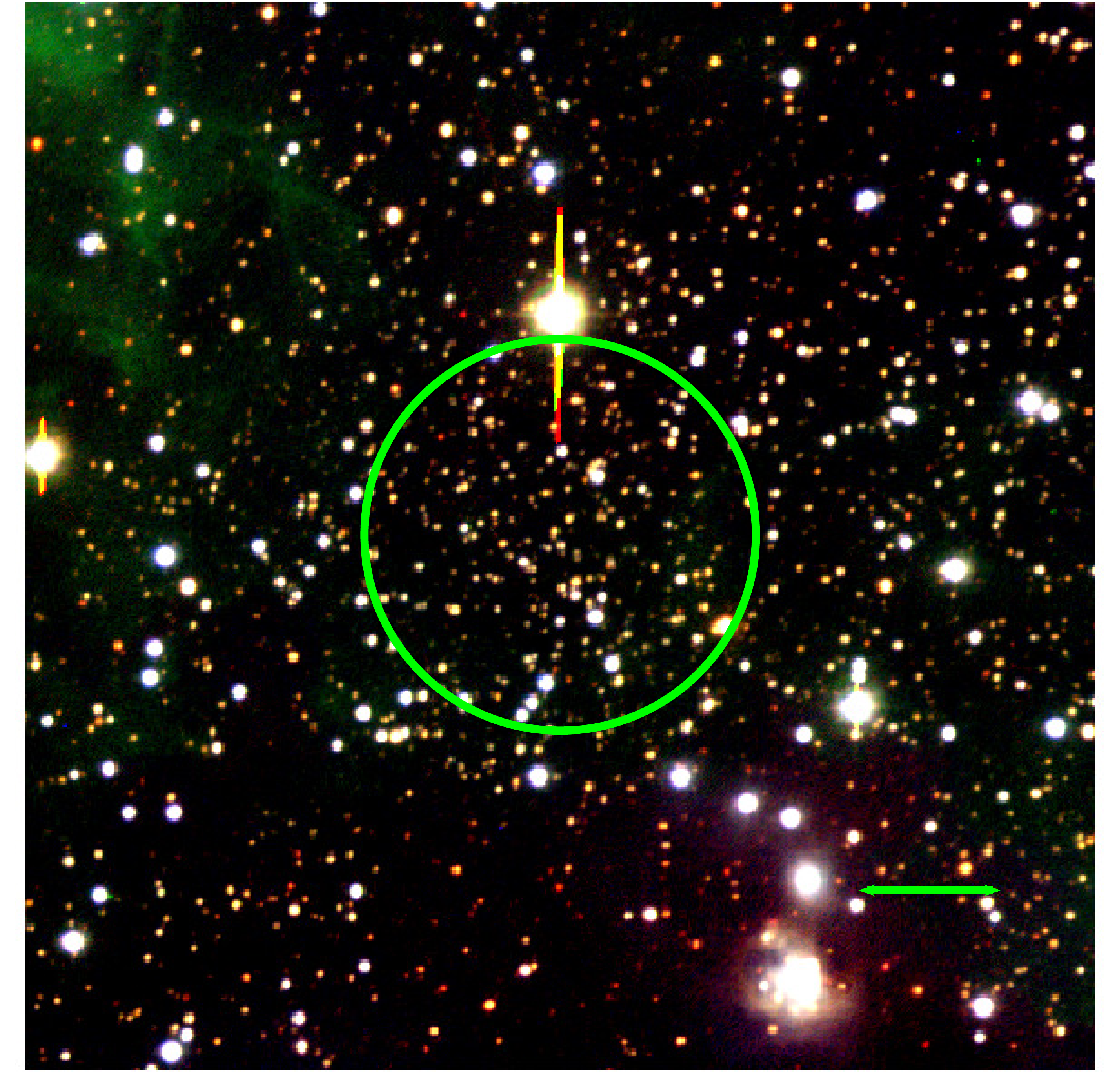}
 \includegraphics[width=80mm]{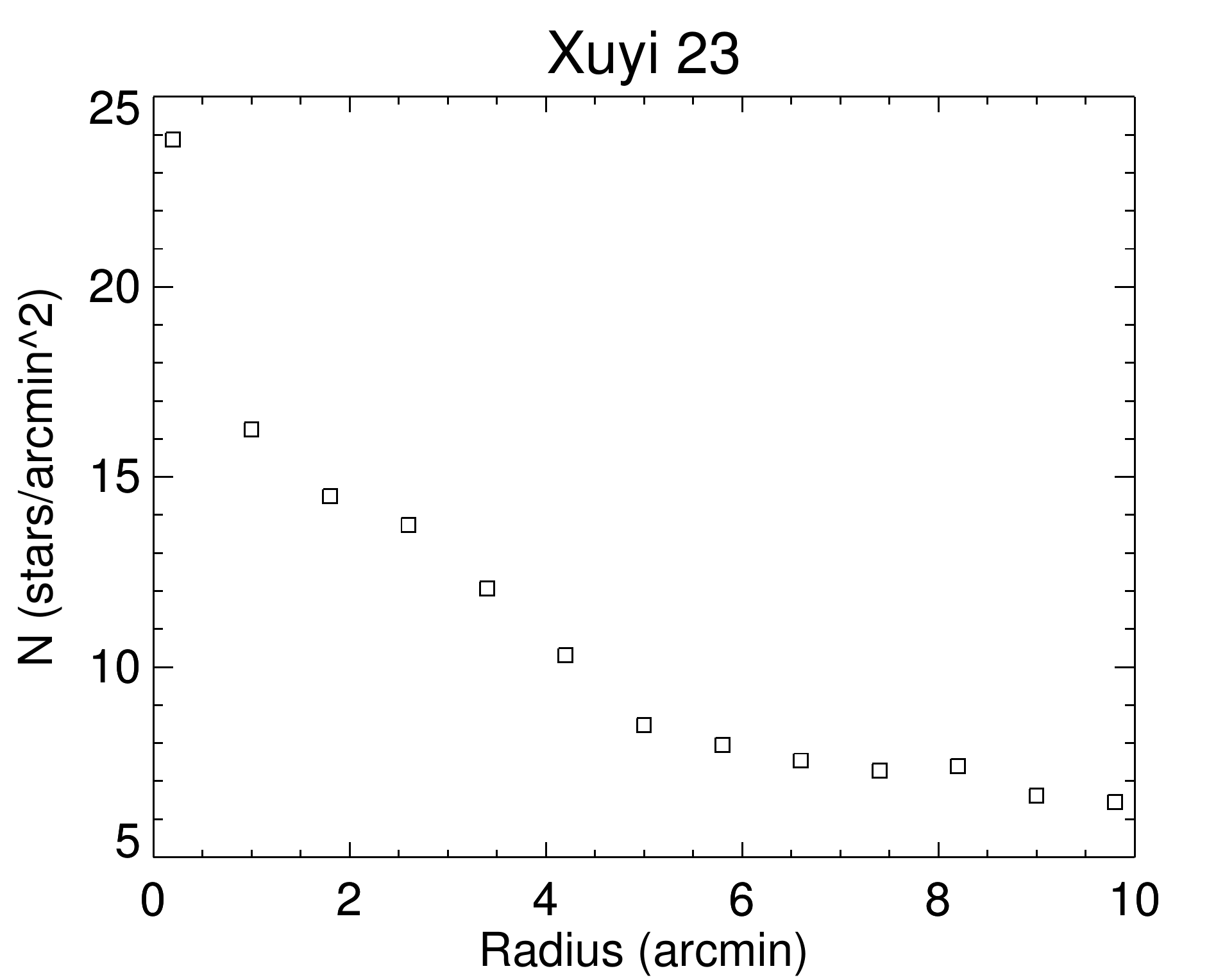}
 \caption{Same as Fig\,\ref{fig1} but for cluster Xuyi 23. Xuyi 23 is one obscured cluster.}
 \label{fig5}
 \end{center}
 \end{figure*}
 
  \begin{figure*}
 \begin{center}
 \includegraphics[width=170mm]{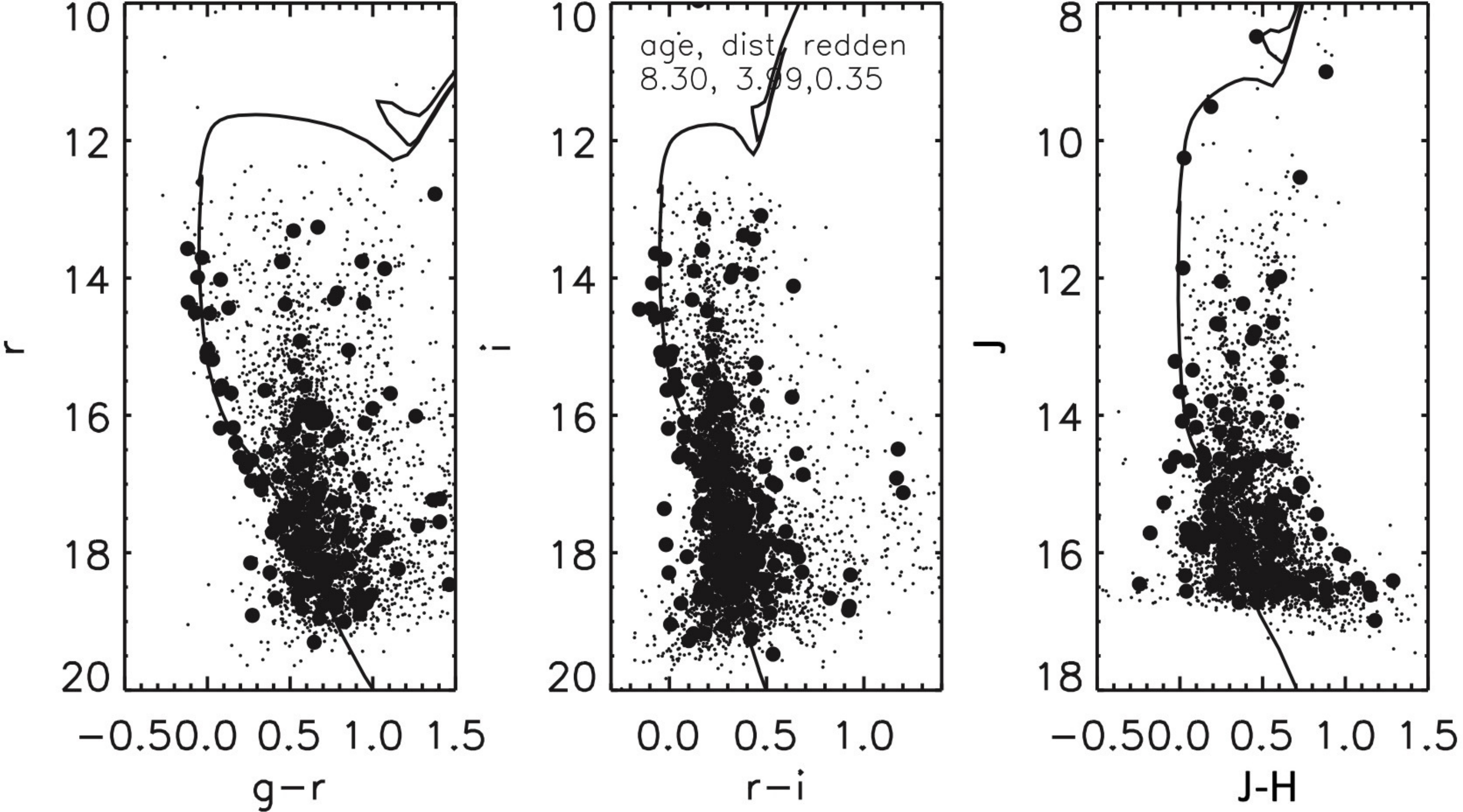}
   \includegraphics[width=60mm]{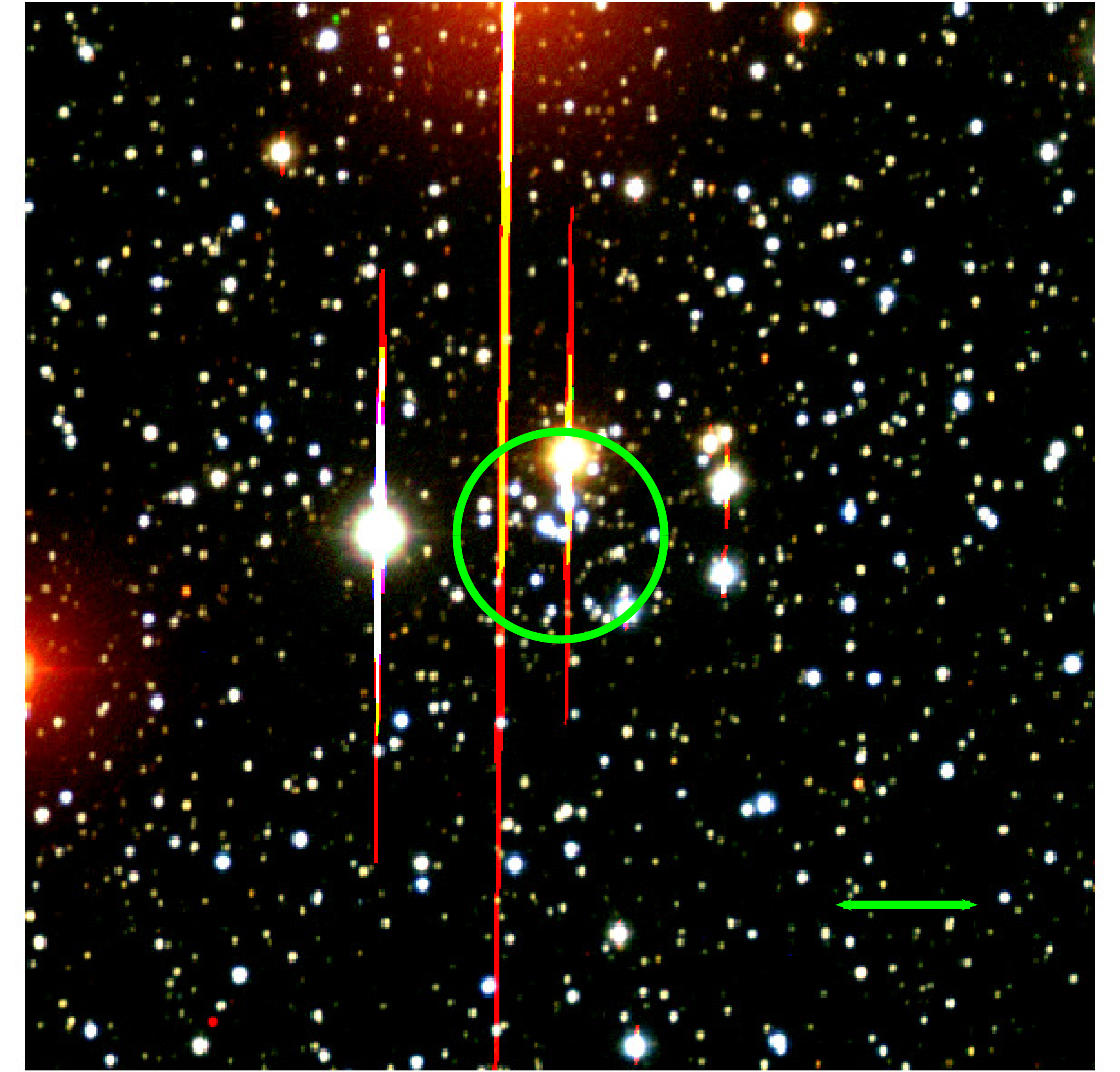}
   \includegraphics[width=80mm]{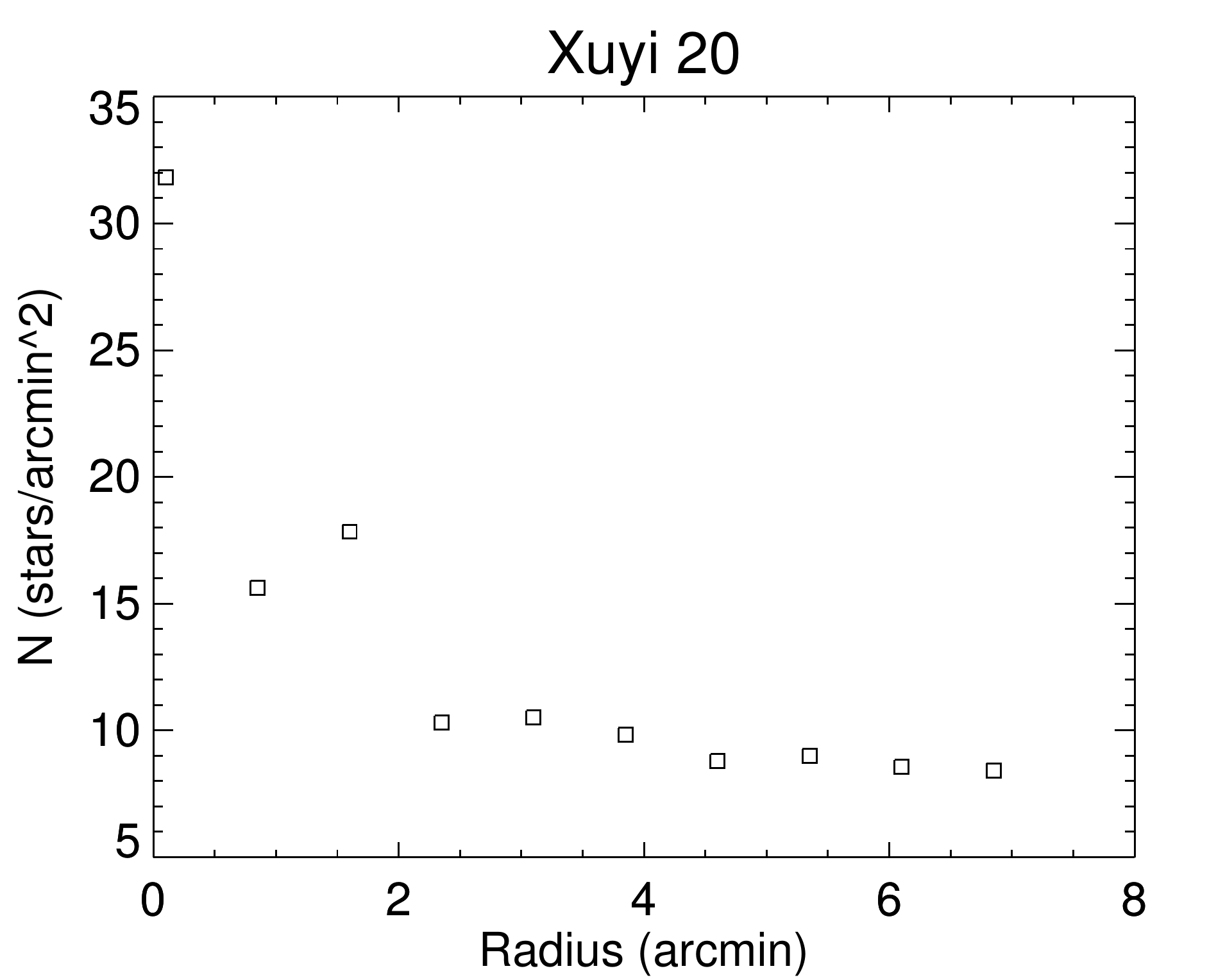}
  \caption{Same as Fig\,\ref{fig1} but for cluster Xuyi 20. Xuyi 20 is one bright cluster.}
 \label{fig6}
 \end{center}
 \end{figure*}

  \begin{figure*}
 \begin{center}
  \includegraphics[width=170mm]{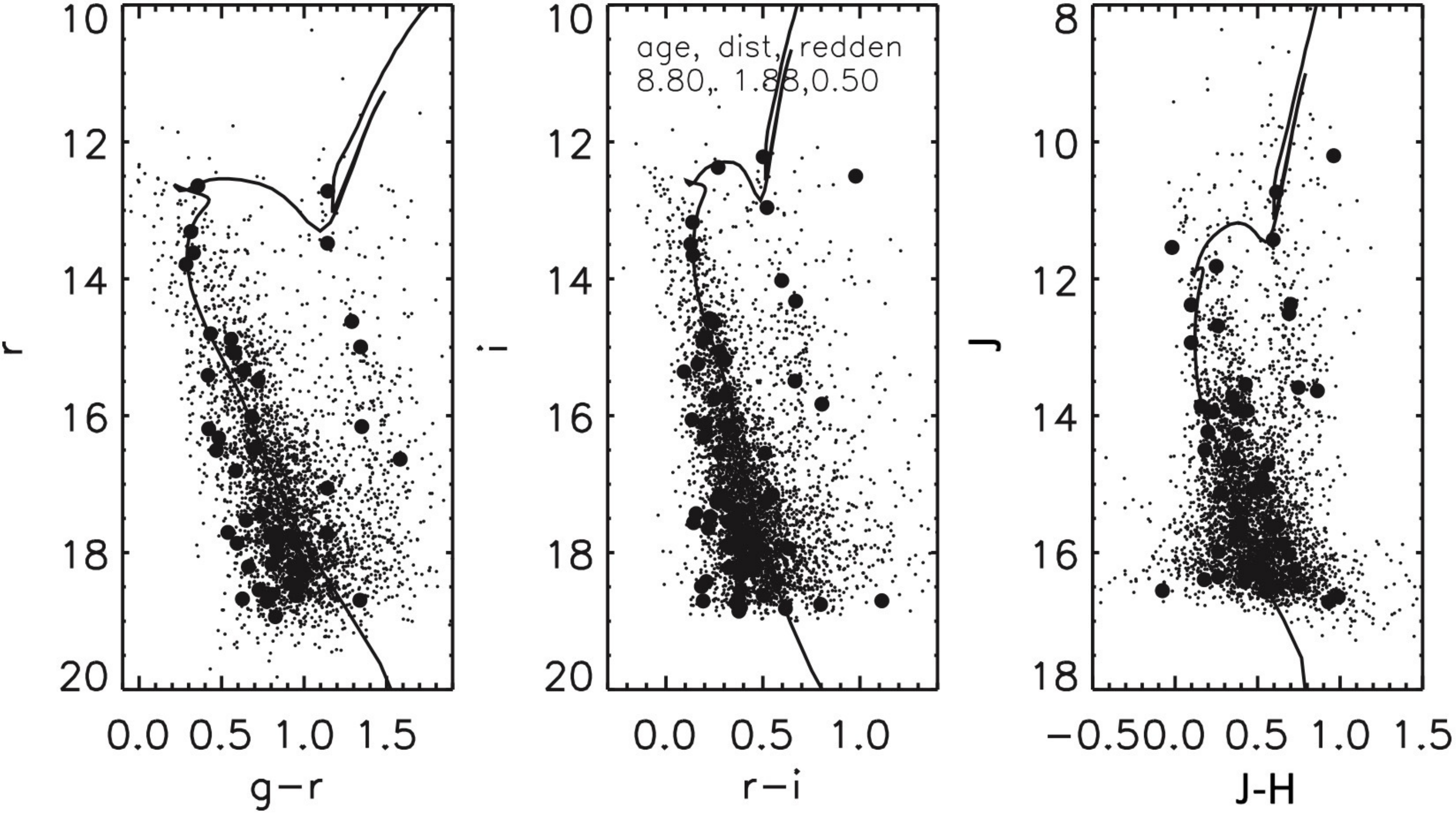}
  \includegraphics[width=60mm]{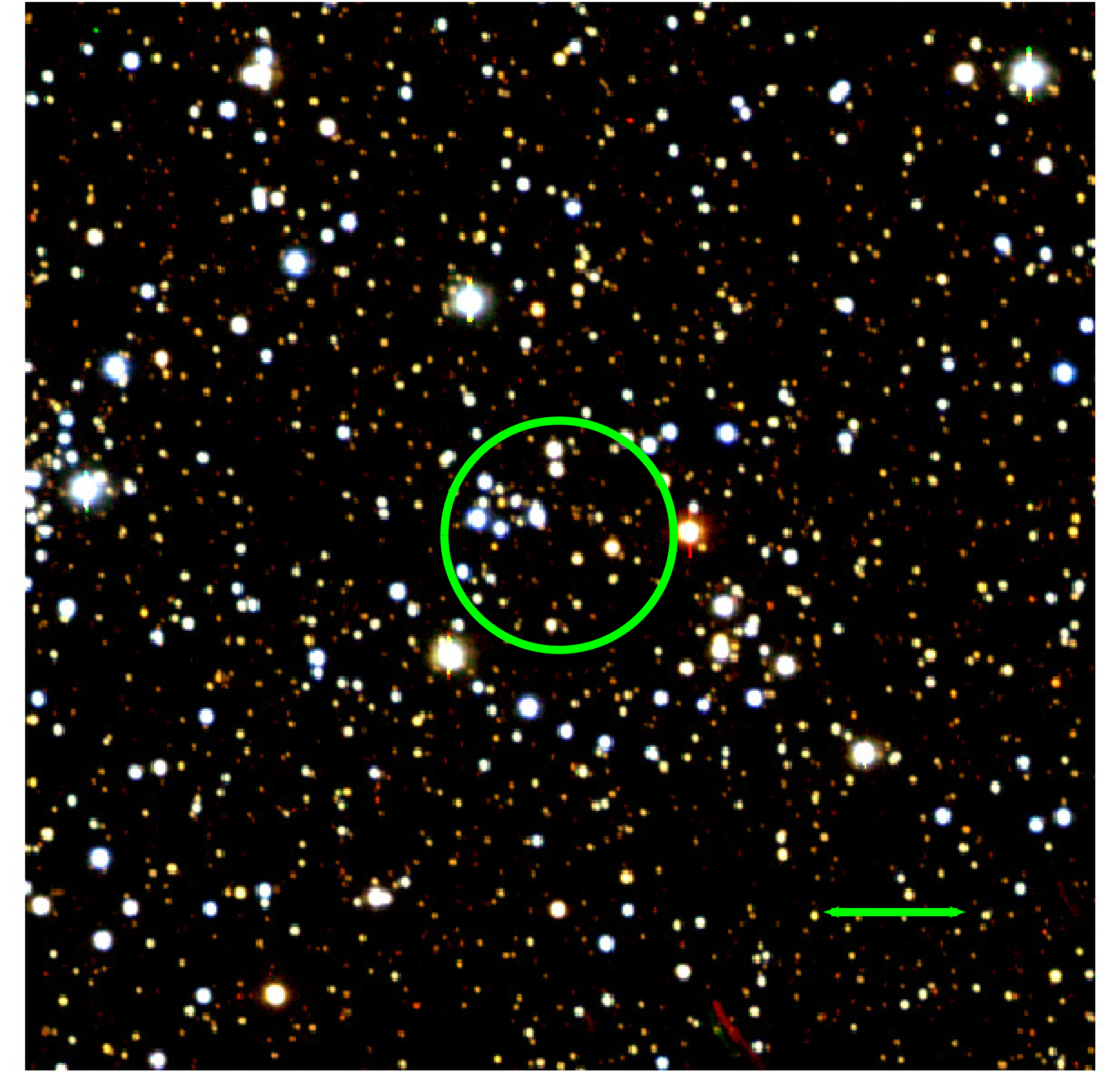}
   \includegraphics[width=80mm]{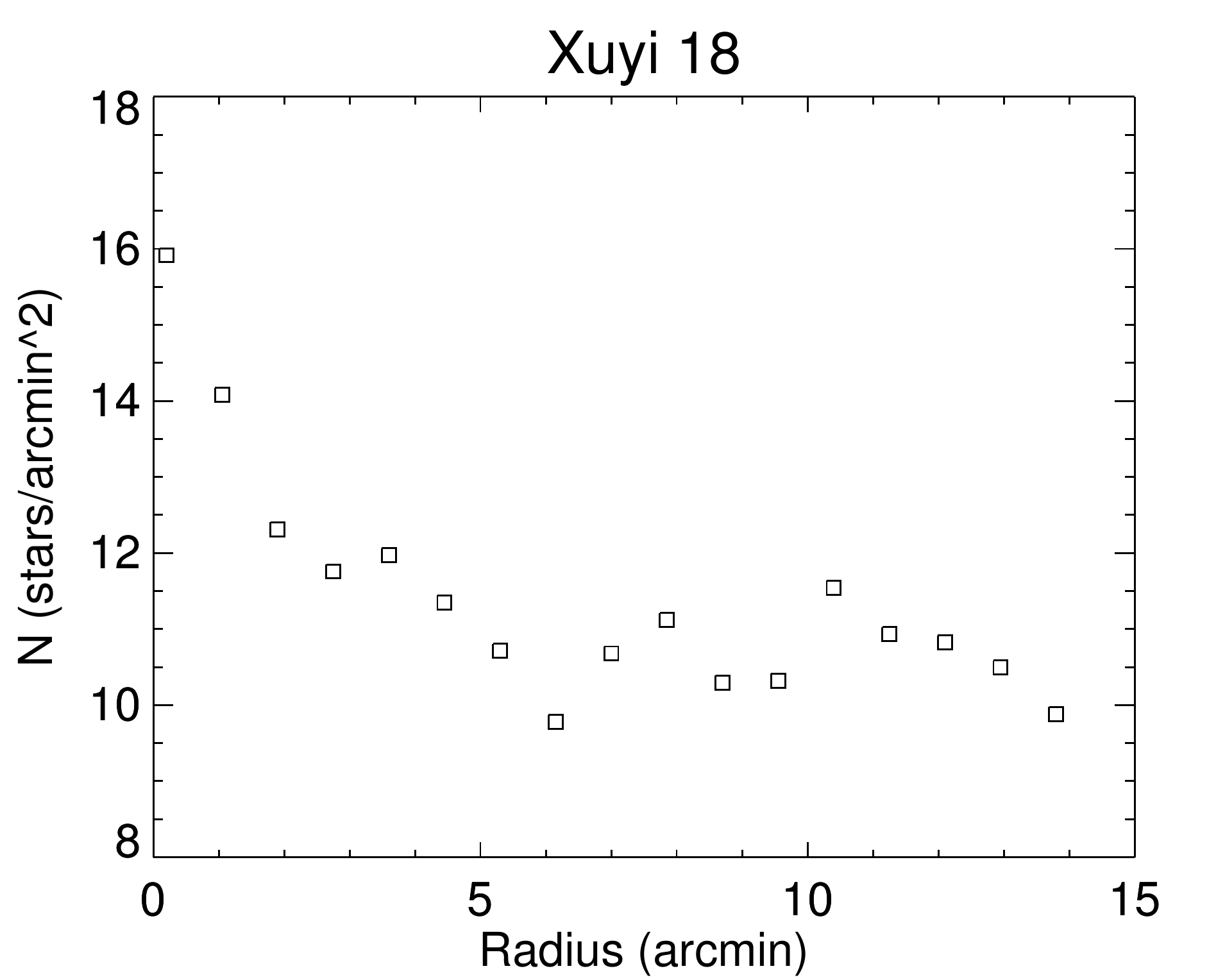} 
  \caption{Same as Fig\,\ref{fig1} but for cluster Xuyi 18.}
 \label{fig8}
 \end{center}
 \end{figure*}
 
 \begin{figure*}
 \begin{center}
 \includegraphics[width=170mm]{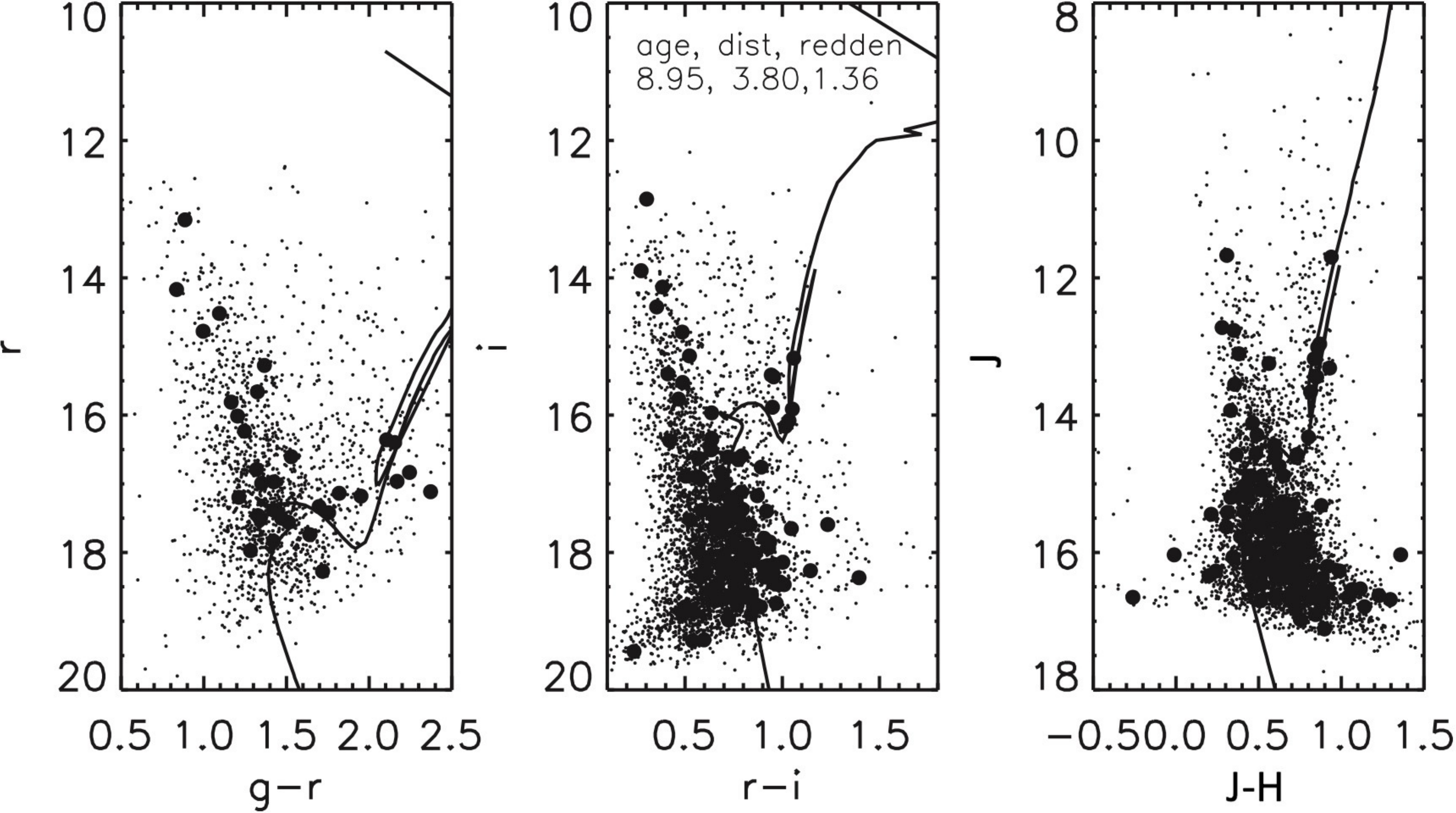}
  \includegraphics[width=60mm]{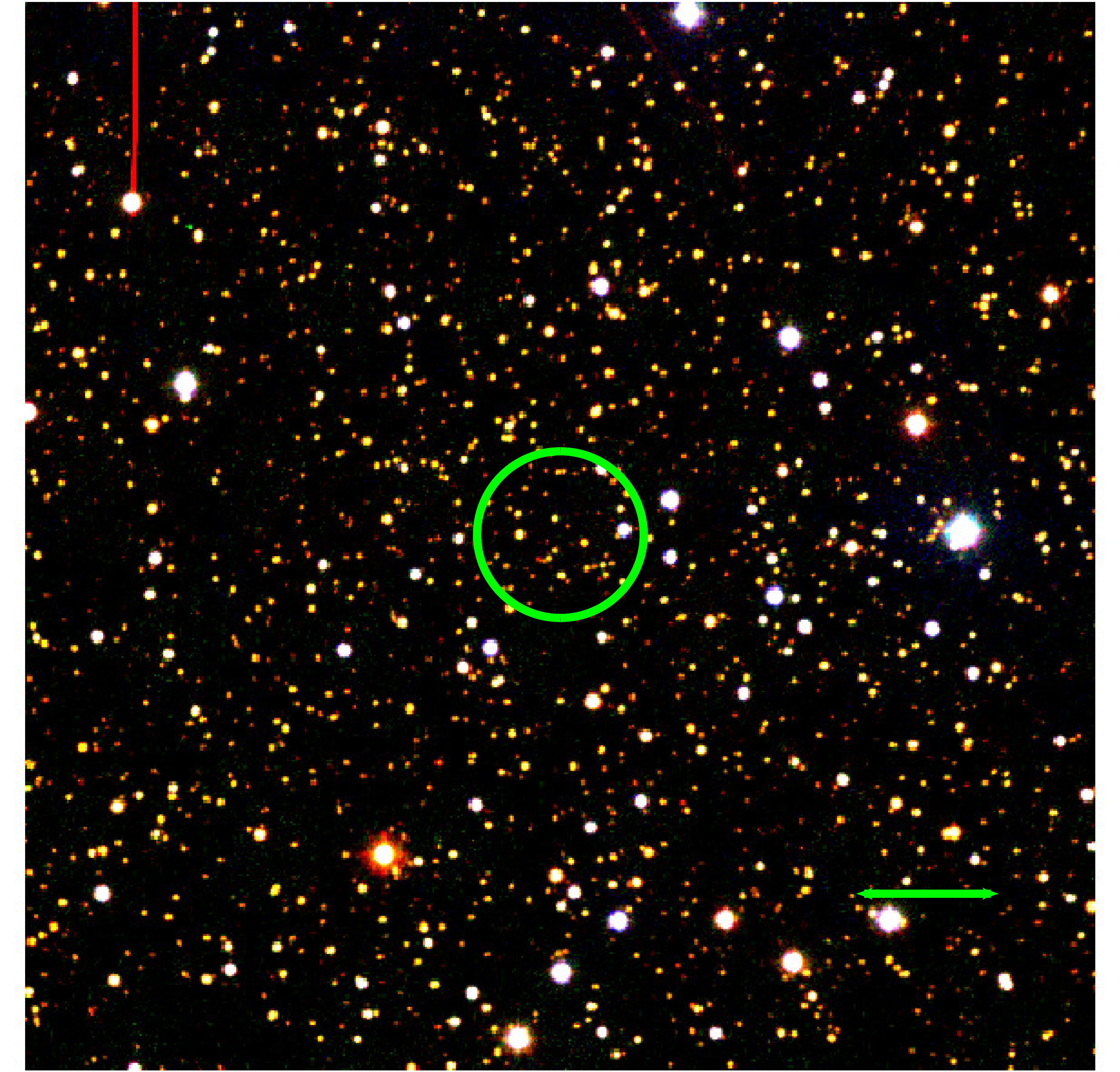}
 \includegraphics[width=80mm]{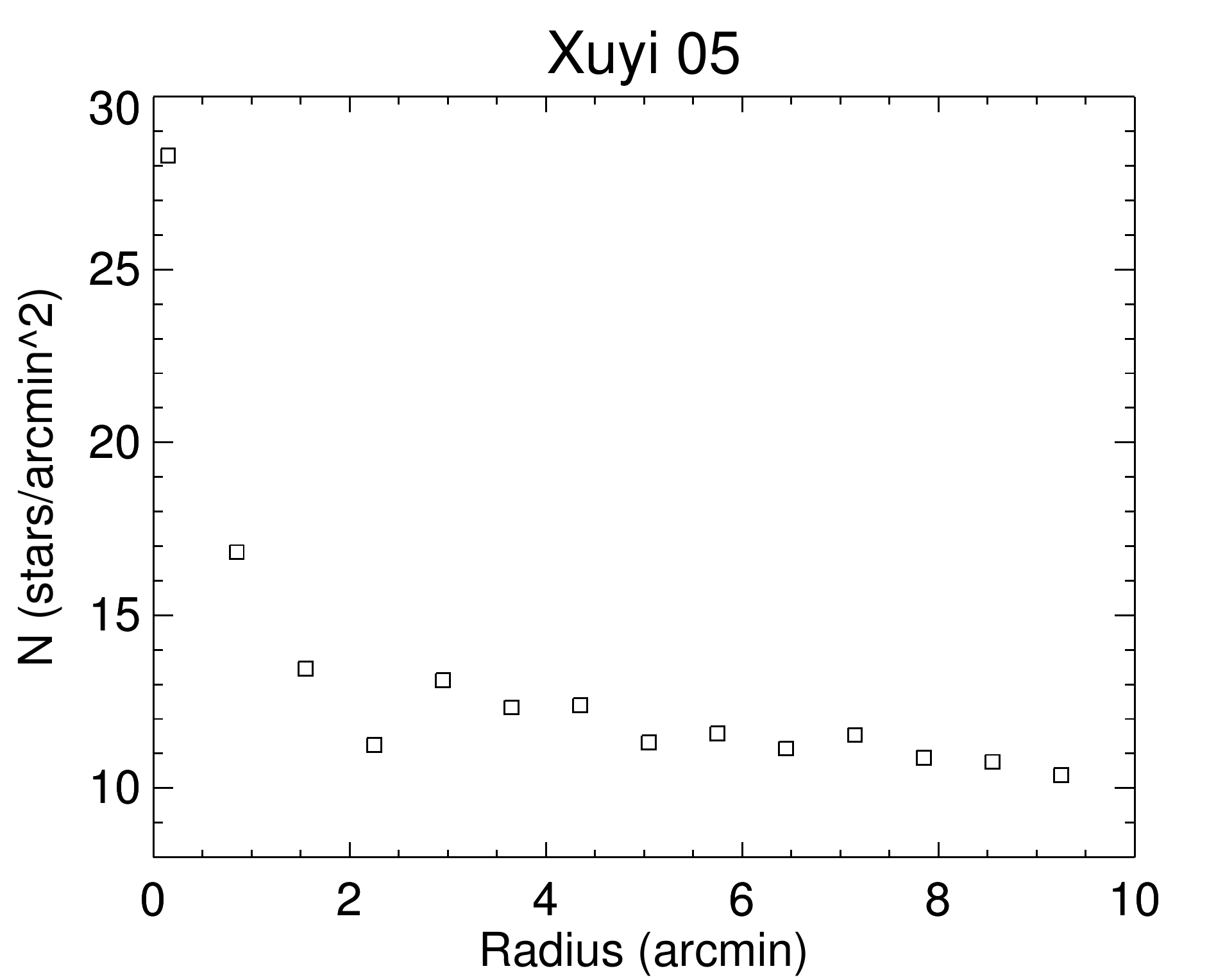}  
 \caption{Same as Fig\,\ref{fig1} but for cluster Xuyi 05. 
 This is one example of inconspicuous clusters that heavily contaminated main sequence is present. 
 Clusters like Xuyi 05 are recognized by some giant stars.}
 \label{fig9}
 \end{center}
 \end{figure*}

 \begin{figure*}
 \begin{center}
 \includegraphics[width=170mm]{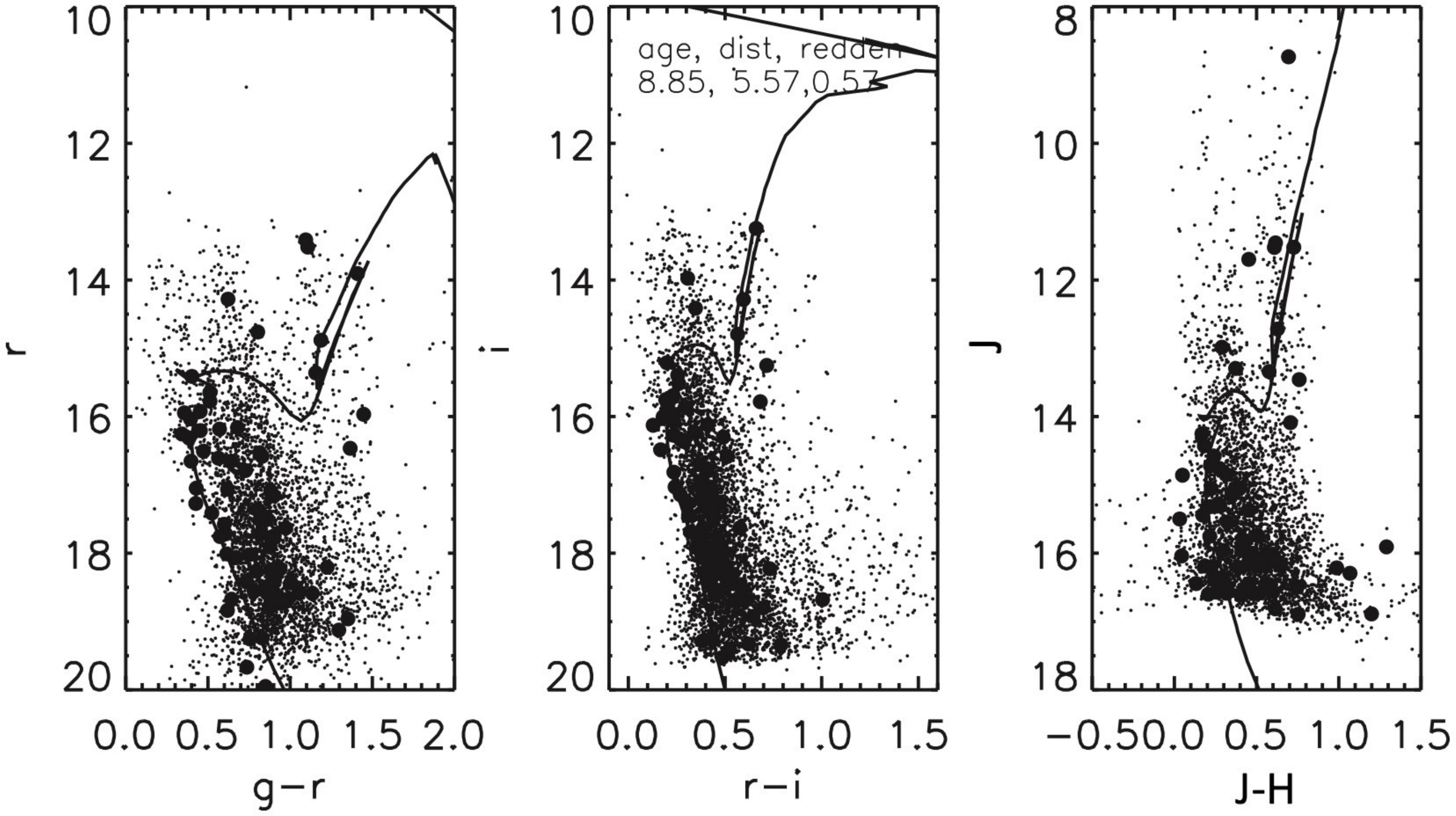}
 \includegraphics[width=60mm]{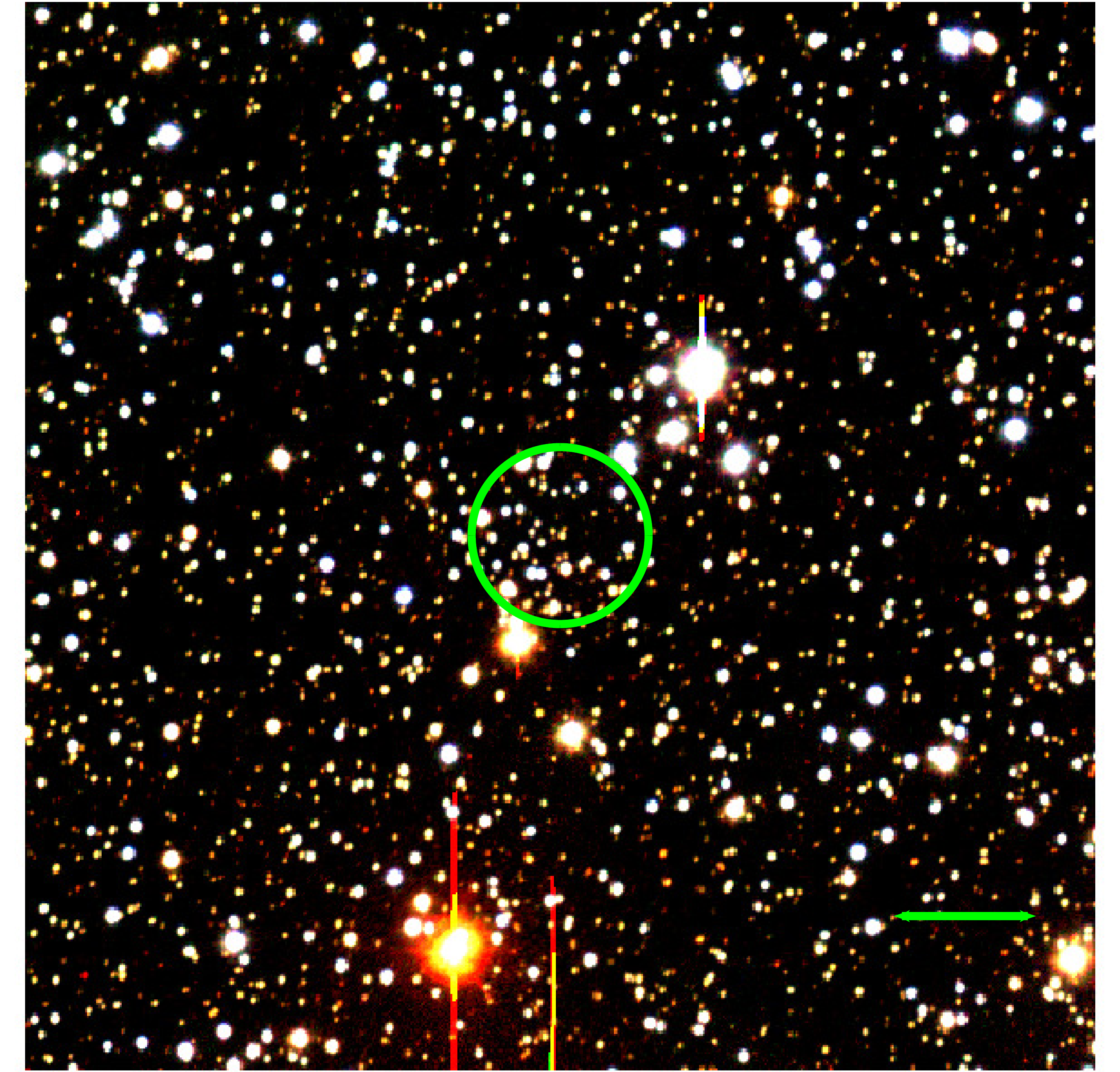}
 \includegraphics[width=80mm]{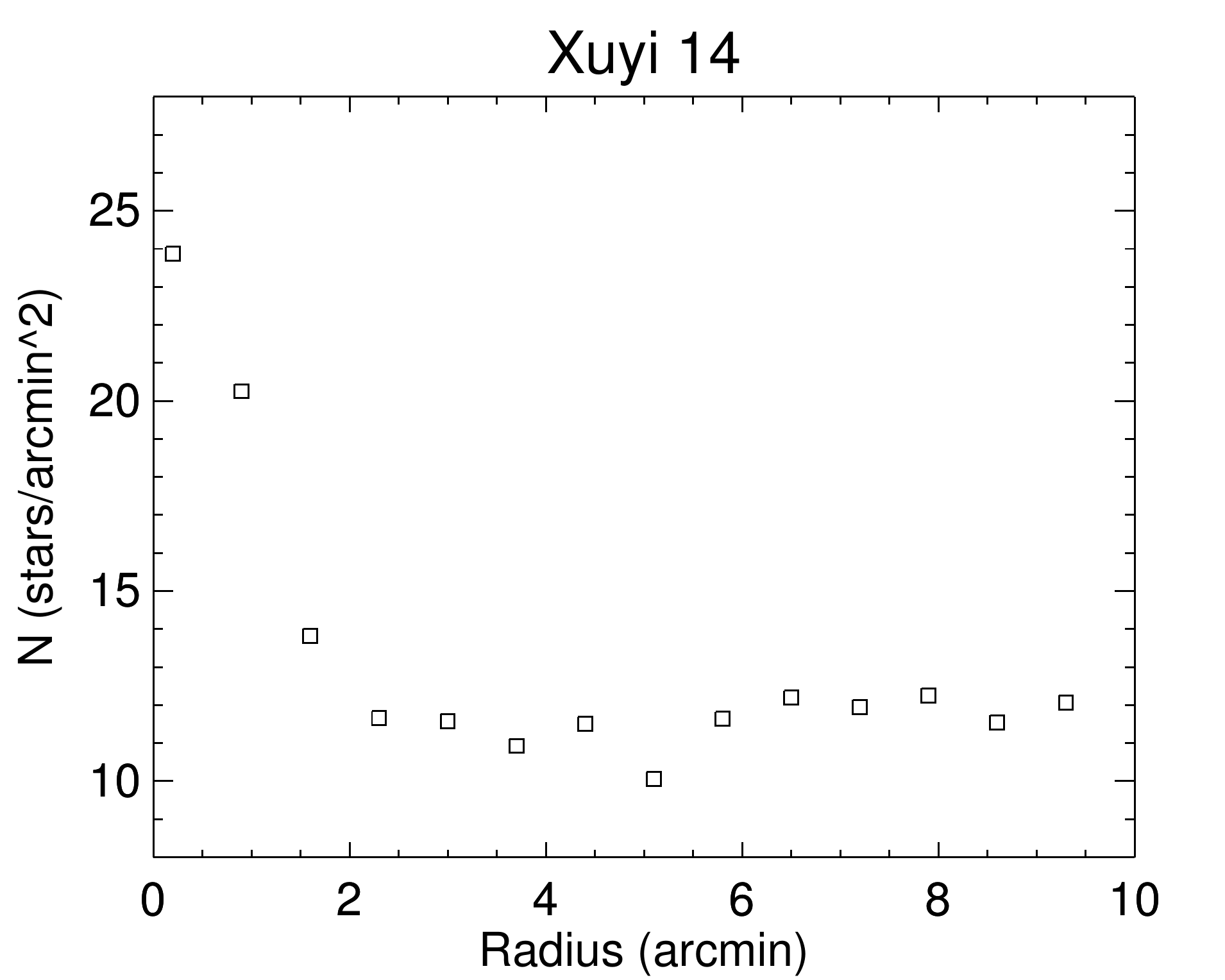} 
 \caption{Same as Fig\,\ref{fig1} but for cluster Xuyi 14. 
 This is another example of inconspicuous clusters that shows heavily contaminated main sequence. 
 Cluster like Xuyi 14 is recognized by some stars on turn-off phase.}
 \label{fig10}
 \end{center}
 \end{figure*}

%----------------------------------------------

%-------------------------------------------------

\end{document}